\documentclass[
twocolumn,
superscriptaddress,
pra,
longbibliography,
showpacs,
preprintnumbers,
amssymb,
floatfix
]{revtex4-1}

\usepackage{amsmath}
\usepackage{bm}
\usepackage{graphicx}
\usepackage[ansinew]{inputenc}
\usepackage{array}
\usepackage{color}
\usepackage{amstext}
\usepackage{amsthm} 
\usepackage{amssymb}
\usepackage{latexsym}
\usepackage{gensymb}
\usepackage{dsfont}
\usepackage{braket}
\usepackage[caption=false]{subfig}

\usepackage[makeroom]{cancel}

\usepackage{balance}
\usepackage{dcolumn}
\usepackage{bm}
\usepackage{bbm}
\usepackage{braket}
\usepackage{mathrsfs}
\usepackage{euscript}
\usepackage{comment}
\usepackage{txfonts}

\usepackage[dvipsnames]{xcolor}

\usepackage[
unicode=true,
bookmarks=false,
breaklinks=false,
pdfborder={0 0 1},
backref=false,
colorlinks=true,
citecolor=blue
]{hyperref}

\usepackage{tikz-cd}

\usepackage{mathrsfs}

\usepackage{amsthm}

\begin{document}

%\preprint{APS/123-QED}

\title{Geometry-Controlled Squeezed-Waveguide Enabled Quantum Batteries}% Force line breaks with \\

\author{Luis D. Zambrano-Palma}
\affiliation{Institute for Quantum Science and Engineering, Texas A\&M University, College Station, Texas 77843, USA}

\author{Yusef Maleki}

 \affiliation{Institute for Quantum Science and Engineering, Texas A\&M University, College Station, Texas 77843, USA}

\author{M. Suhail Zubairy}

\affiliation{Institute for Quantum Science and Engineering, Texas A\&M University, College Station, Texas 77843, USA}

\date{\today}

\begin{abstract}
Harnessed nonclassical resources to enhance
the performance of quantum batteries is an important goal of quantum technology. In this work, we investigate
multi-emitter batteries coupled to a one-dimensional waveguide
driven by a broadband squeezed-vacuum reservoir, and determine how the
waveguide geometry controls the battery performance and 
and their thermodynamic features. We show that for a single emitter,
the squeezed reservoir deposits finite energy while the state remains
passive, yielding zero ergotropy. For two emitters, however,
phase-sensitive reservoir correlations generate coherence and can drive
the steady state into a fully nonpassive regime. For three emitters,
collective thermal dynamics already supports finite ergotropy, while
squeezing further increases the extractable component of the stored
energy. Crucially, because the collective and phase-sensitive couplings
depend on the emitter positions, geometry provides direct control over
the nonclassical correlations and, consequently, over the nonpassivity
of the steady state. The ergotropy can therefore be strongly
modulated while the stored energy changes only weakly, and in the
two-emitter configuration geometry alone can switch the battery between
a passive steady state and one in which all stored energy is unitarily
extractable. %Consistently, the ergotropy maxima closely track the coherence peaks, whereas pairwise entanglement exhibits a more complex spatial redistribution.
\end{abstract}

\maketitle

\section{Introduction}

Quantum technologies rely on the preparation, manipulation, and
characterization of quantum resources that enable operational
capabilities beyond those accessible to classical systems. Among the
most widely studied are quantum coherence, entanglement, and quantum
squeezing, which play central roles in the processing, transmission, and
control of quantum information. These resources underpin advances in
quantum computing~\cite{ladd2010quantum,nielsen2010quantum},
quantum communication~\cite{usenko2026continuous,hasan2023quantum},
quantum simulation~\cite{maleki2023time,halimeh2025cold,daley2022practical}, quantum memories~\cite{fleischhauer2002quantum,lvovsky2009optical}, 
quantum illumination and radar protocols~\cite{karsa2024quantum,palma2026optimal},
quantum metrology~\cite{maleki2022distributed, pirandola2018advances,huang2024entanglement},
and quantum measurement and control
~\cite{4h4b-3xss,doi:10.1142/S0219477525400280}. Identifying which resource is operationally relevant in a given
setting, however, can be nontrivial, particularly for open or mixed
quantum systems in which different forms of nonclassicality need not
contribute in the same way to the physical task under consideration.

A clear setting in which these questions become directly relevant is
provided by Quantum batteries (QBs)~\cite{campaioli2024colloquium,ferraro2026opportunities}. In
their simplest realization, a QB can be composed of one
or several quantum cells, each of which may be modeled as a two-level
system whose ground and excited states define the available energy
levels. Charging then corresponds to driving the battery away from its
initial low-energy state by populating higher-energy levels and, more
generally, by generating quantum coherences and correlations within the
system. However, the total energy stored in a QB does not necessarily coincide with the amount of useful work
that can be extracted from it, since a state may possess finite
internal energy while remaining passive~\cite{pusz1978passive}. For this purpose, the relevant quantity is the ergotropy~\cite{allahverdyan2004maximal}, which measures the maximum work extractable
through cyclic unitary operations. From this perspective, the central
problem is not only to increase the energy of the QB, but also to
control the structure of its quantum state so that a large fraction of
that energy remains extractable. This naturally connects
QB performance with nonclassical resources such as quantum entanglement~\cite{kamin2020entanglement,alicki2013entanglement,hovhannisyan2013entanglement}, coherence~\cite{andolina2019extractable,centrone2021charging,li2025enhancing} and many-body correlations~\cite{rossini2019many,binder2015quantacell}, which can alter the nonpassivity
of the QB state and thereby influence its thermodynamic
usefulness. 

On the other hand, light--matter interfaces provide a natural platform
for controlling nonclassical resources and engineering the quantum
states of few- and many-body systems. Cavity quantum
electrodynamics (QED), and its circuit-QED counterpart, have long
provided a means of enhancing and manipulating the interaction between
quantum emitters and confined electromagnetic fields
~\cite{RevModPhys.93.025005}. Waveguide QED extends this paradigm by
coupling localized quantum emitters to a continuum of propagating
photonic modes in a quasi-one-dimensional geometry
~\cite{RevModPhys.89.021001,RevModPhys.95.015002,zhou2023emission}. The resulting interactions can be long ranged and strongly dependent
on the relative positions of the emitters, giving rise to collective
phenomena such as superradiance and subradiance, as well as correlated
photon transport and emitter--emitter correlations. These properties have motivated the use of waveguide-QED platforms for
the generation, manipulation, and detection of entanglement
~\cite{PhysRevA.89.042328}, as well as for photonic quantum
information processing based on propagating photons
~\cite{PhysRevLett.111.090502}.

The properties of such systems can be further enriched by engineering
the electromagnetic environment itself. Squeezed-vacuum reservoirs,
for example, introduce phase-sensitive two-photon correlations that
lead to dynamics absent in ordinary vacuum or thermal environments and
have been shown experimentally to modify radiative properties such as
resonance fluorescence and dephasing
~\cite{PhysRevX.6.031004}. In one-dimensional squeezed waveguides, the
ordinary collective interaction depends on the relative emitter
separation, whereas the phase-sensitive processes depend on the
emitter positions relative to the squeezing sources
~\cite{you2018waveguide}. This spatial dependence enables geometric
control of decay, dephasing, and steady state entanglement, and naturally
raises the question of whether the same mechanism can be used in QBs
to control not only stored energy but also its extractable fraction.

This question is closely connected to the broader use of engineered
environments in open quantum batteries, where the reservoir can modify
both the charging dynamics and the fraction of stored energy that
remains extractable. Thermal reservoirs, for example, can sustain
finite steady-state ergotropy when several qubits are collectively
coupled to the bath~\cite{PhysRevE.102.042111}, while squeezed thermal
environments can enhance charging through coherence generation
~\cite{centrone2021charging}. More recent reservoir-engineering
approaches have further shown that squeezed reservoirs can enhance both
stored energy and ergotropy under suitable conditions
~\cite{b5lx-j66h}. Waveguide environments provide an additional degree
of control by structuring the photonic modes that mediate the
interaction between the QB and its surroundings. They have been
used to enable remote charging and suppress degradation through
environment-induced bound states~\cite{PhysRevLett.132.090401}, while
collective and interference effects in waveguide-QED architectures can
prolong the storage of energy and ergotropy or enhance energy transfer
~\cite{tirone2025many,PhysRevLett.134.180401,43n6-rnj3,h3tl-lxt8}.
These results show that both the quantum state of the reservoir and the
spatial structure of the electromagnetic environment can play a
critical role in determining the thermodynamic performance of a QB.

Motivated by these developments, we consider a QB in which a broadband
squeezed-vacuum field acts as the charger for quantum emitters coupled
to a one-dimensional waveguide. We show that a single emitter can
store finite energy while remaining passive, whereas for two emitters
the squeezed-reservoir correlations generate coherence and can drive
the steady state into a fully nonpassive regime. For suitable geometry,
the entire stored energy becomes unitarily extractable, while a thermal
reservoir with the same mean occupation yields zero ergotropy. For
three emitters, collective thermal dynamics already produces finite
ergotropy, which is further enhanced by squeezing. Most importantly,
the position dependence of the waveguide couplings enables geometric
control of nonpassivity, allowing the ergotropy to vary strongly while
the stored energy changes only weakly. We further find that this
geometry-dependent ergotropy closely follows the generated coherence,
whereas pairwise entanglement does not exhibit a direct monotonic
relation with extractable work.

This work is organized as follows.
Section~\ref{sec2} introduces the squeezed-waveguide model and the
energetic quantities used to characterize the QB.
Section~\ref{sec3} examines the charging dynamics of one-, two-, and
three-emitter batteries and analyzes the emergence of finite ergotropy
under collective coupling. Section~\ref{sec4} investigates the
transient dynamics and establishes how emitter geometry and squeezing
control the stored energy and extractable work. Section~\ref{sec5}
examines the roles of coherence and entanglement and their
relation to the geometry-dependent ergotropy. Conclusions are
presented in Section~\ref{sec6}.

\section{Model and energetic quantities}
\label{sec2}

\begin{figure*}[!htbp]
    \centering \includegraphics[width=\textwidth]{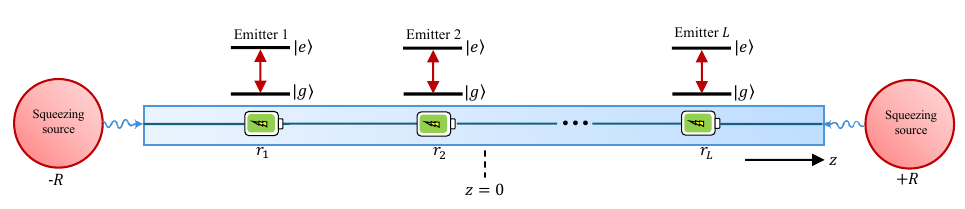}
\caption{
Schematic of the squeezed-waveguide quantum battery. The battery is
formed by $L$ two-level emitters located at positions $r_i$ along a
one-dimensional waveguide. Broadband squeezed-vacuum fields generated
by squeezing sources located at $\pm R$ propagate through the
waveguide and act as the charger. The coordinate origin is chosen at
the midpoint between the two squeezing sources.
}\label{fig:setup}
\end{figure*}

In this section, we introduce the physical model of the QB, illustrated
schematically in Fig.~\ref{fig:setup}. The battery consists of $L$
identical two-level emitters coupled to a one-dimensional waveguide,
while broadband squeezed-vacuum fields generated by two squeezing
sources and propagating through the waveguide act as the charger. The total emitter--field Hamiltonian
is written as~\cite{you2018waveguide}
\begin{equation}
H=H_A+H_F+H_{AF},
\label{eq:total_H}
\end{equation}
where $H_A=
\sum_{i=1}^{L}
\hbar\omega_0
|e_i\rangle\langle e_i|$ is the atomic Hamiltonian. Each emitter has a ground state
$|g_i\rangle$, an excited state $|e_i\rangle$, and the same transition
frequency $\omega_0$. The electromagnetic field Hamiltonian is $H_F=
\sum_{\mathbf{k},s}
\hbar\omega_{\mathbf{k}s}
\left(
\hat a_{\mathbf{k}s}^{\dagger}
\hat a_{\mathbf{k}s}
+\frac{1}{2}
\right),$ while the emitter--field interaction in the electric-dipole
approximation is $H_{AF}
=
-i\hbar
\sum_{\mathbf{k},s}
\sum_{i=1}^{L}
\Big[\boldsymbol{\mu}_i\cdot
\mathbf{u}_{\mathbf{k}s}(\mathbf{r}_i)
S_i^{+}\hat a_{\mathbf{k}s}+
\boldsymbol{\mu}_i^{*}\cdot
\mathbf{u}_{\mathbf{k}s}(\mathbf{r}_i)
S_i^{-}\hat a_{\mathbf{k}s}
-\mathrm{H.c.}
\Big],$ where $S_i^{+}=|e_i\rangle\langle g_i|$ and
$S_i^{-}=|g_i\rangle\langle e_i|$ are the raising and lowering
operators of the $i$th emitter.

After we have introduced the microscopic emitter--field Hamiltonian, we now
consider the reduced open-system dynamics of the QB.
The emitters interact with the electromagnetic modes supported by the
one-dimensional waveguide, which are prepared in a broadband
squeezed-vacuum state and constitute the photonic reservoir. By tracing
over these field degrees of freedom, the dynamics is reduced to the
density operator $\rho$ of the $L$-emitter system. The guided-mode
structure determines the coherent and dissipative couplings between
the emitters and introduces an explicit dependence on their positions
along the waveguide. Consequently, both the ordinary radiative
processes and the phase-sensitive squeezed-vacuum contributions are
controlled by the emitter geometry. The reduced dynamics is governed
by the master equation \cite{you2018waveguide}
\begin{align}
\frac{d\rho}{dt}
={}&
-i\sum_{i\neq j}
\Omega_{ij}
\left[
S_i^{+}S_j^{-},\rho
\right]e^{ i(\omega_i-\omega_j)t}
\nonumber\\
&-\frac{1}{2}
\sum_{i,j}
\gamma_{ij}(1+N)
\left(
\rho S_i^{+}S_j^{-}
+
S_i^{+}S_j^{-}\rho
-
2S_j^{-}\rho S_i^{+}
\right)e^{ i(\omega_i-\omega_j)t}
\nonumber\\
&-\frac{1}{2}
\sum_{i,j}
\gamma_{ij}N
\left(
\rho S_i^{-}S_j^{+}
+
S_i^{-}S_j^{+}\rho
-
2S_j^{+}\rho S_i^{-}
\right)e^{ -i(\omega_i-\omega_j)t}
\nonumber\\
&-\frac{1}{2}
\sum_{\alpha=\pm}
\sum_{i,j}
\gamma_{ij}^{\prime}
M e^{2\alpha i k_{0z}R}
\left(
\rho S_i^{\alpha}S_j^{\alpha}
+
S_i^{\alpha}S_j^{\alpha}\rho
-
2S_j^{\alpha}\rho S_i^{\alpha}
\right),
\label{eqmaster}
\end{align}
where $N=\sinh^2(r)$ is the mean photon number of the squeezed-vacuum
reservoir, $M=\sinh(r)\cosh(r)=\sqrt{N(N+1)}$ is the two-photon
correlation amplitude, and $r$ denotes the squeezing parameter.

Here, $\Omega_{ij}$ describes the coherent dipole--dipole interaction,
$\gamma_{ij}$ denotes the collective radiative decay rate associated
with the ordinary vacuum contribution, and $\gamma_{ij}^{\prime}$
characterizes the phase-sensitive collective two-photon processes
induced by the squeezed reservoir. For identical emitters,
$\omega_i=\omega_j=\omega_0$, so the phase factors associated with the
frequency differences in Eq.~\eqref{eqmaster} reduce to unity. For the
one-dimensional waveguide considered here, these coefficients take the
form~\cite{you2018waveguide}
\begin{subequations}
\label{eq:1D_coefficients}
\begin{align}
\gamma_{ij}
&=
\gamma_{1d}
\cos(k_{0z}r_{ij}), \label{gammaij}
\\
\Omega_{ij}
&=
\frac{\gamma_{1d}}{2}
\sin(k_{0z}r_{ij}), \label{omegaij}
\\
\gamma_{ij}^{\prime}
&=
\gamma_{1d}
\cos\left[
k_{0z}(r_i+r_j)
\right]. \label{gammapij}
\end{align}
\end{subequations}
Here, $\gamma_{1d}$ is the single-emitter decay rate into the guided modes,
$r_{ij}=\lvert r_i-r_j\rvert$ is the separation between emitters $i$
and $j$, and $\lambda_{0z}=2\pi/k_{0z}$ is the effective longitudinal
wavelength of the waveguide. The coordinate origin is chosen at the
midpoint between the two squeezing sources, located symmetrically at
$\pm R$. %It follows from Eq.~\eqref{eq:1D_coefficients} that $\gamma_{ij}$ and $\Omega_{ij}$ depend on the relative emitter separation, whereas $\gamma_{ij}^{\prime}$ depends on $r_i+r_j$ and therefore on the emitter positions relative to the squeezing sources.

Once the reduced dynamics has been specified, we introduce the energetic
quantities used to characterize the charging performance of the quantum
battery. A central quantity is the energy stored in the $L$-emitter battery
at time $t$, defined as
\begin{equation}
E_L(t)
=
\operatorname{Tr}
\left[
\rho(t)H_B
\right],
\label{eq:energy}
\end{equation}
where
$H_B=\hbar\omega_0\sum_{i=1}^{L}S_i^{+}S_i^{-}$
is the bare battery Hamiltonian, whose spectrum extends from $0$ to
$L\hbar\omega_0$. Throughout this work, the battery is initially prepared in
the collective ground state,
$\rho(0)=|G\rangle\langle G|$, with
$|G\rangle=|g\rangle^{\otimes L}$. Since this state has zero energy with
respect to $H_B$, Eq.~\eqref{eq:energy} directly quantifies the energy
accumulated in the battery during the charging process.

The stored energy alone, however, does not determine how much work can
be extracted from the battery. The maximum work extractable through a
cyclic unitary transformation is quantified by the ergotropy,
\begin{equation}
\mathcal{W}_L(t)
=
\operatorname{Tr}\!\left[\rho(t)H_B\right]
-
\min_U
\operatorname{Tr}
\left[
U\rho(t)U^{\dagger}H_B
\right].
\label{eq:ergotropy_def}
\end{equation}
The minimizing unitary transforms $\rho$ into its associated passive
state. Writing the spectral decomposition of the battery state as $\rho
=
\sum_{n=1}^{2^L}
r_n
|r_n\rangle
\langle r_n|,$ where $r_1\ge r_2\ge\cdots\ge r_{2^L},$ and denoting the eigenstates of $H_B$ by $|\epsilon_n\rangle$, with
energies ordered as $\epsilon_1\le\epsilon_2\le\cdots\le\epsilon_{2^L},$ the passive state is obtained by assigning the largest eigenvalues of
$\rho$ to the lowest energy levels, $\rho_{\mathrm{pas}}
=
\sum_{n=1}^{2^L}
r_n
|\epsilon_n\rangle
\langle\epsilon_n|.$ Its energy is therefore $E_{\mathrm{pas},L}(t)
=\sum_{n=1}^{2^L}
r_n(t)\epsilon_n,$ so that the ergotropy can equivalently be written as~\cite{allahverdyan2004maximal,hatsopoulos1976unified,campaioli2024colloquium}
\begin{equation}
\mathcal{W}_L(t)
=
E_L(t)-E_{\mathrm{pas},L}(t).
\label{eq:ergotropy}
\end{equation}
From the definitions above, a passive state satisfies
$\mathcal{W}_L=0$, even when its stored energy is finite. In this case,
the deposited energy cannot be extracted through cyclic unitary
operations and therefore does not constitute useful work storage. By
contrast, a nonpassive state has $\mathcal{W}_L>0$, indicating that at
least part of the stored energy is extractable. In the limiting case
$\mathcal{W}_L=E_L$, the entire stored energy can be extracted by a
suitable unitary transformation.

\section{Few-emitter QB dynamics}
\label{sec3}

\subsection{Single-emitter battery}
\label{sec:single}

In this subsection, we first examine the charging performance of the
single-emitter QB ($L=1$), initially prepared in the ground
state, $\rho_1(0)=|g\rangle\langle g|$. For a single emitter located at $r_1=\delta$ along the waveguide,
the relevant coupling parameters reduce to
$\gamma=\gamma_{1d}$ and
$\gamma'=\gamma_{1d}\cos(2k_{0z}\delta)$. In the energy basis $\{|e\rangle,|g\rangle\}$, the resulting
time-dependent density matrix is given by
\begin{equation}
\rho_{1}(t)
=
\begin{pmatrix}
p(t) & 0\\
0 & 1-p(t)
\end{pmatrix},
\label{DM1}
\end{equation}

with

\begin{equation}
p(t)
=
\frac{N}{2N+1}
\left[
1-e^{-\gamma(2N+1)t}
\right].
\end{equation}
Using the density matrix above, the energy stored in the atomic system at time $t$ follows directly from
\begin{equation}
E_1(t)
=
\hbar\omega_0
\frac{N}{2N+1}
\left[
1-e^{-\gamma(2N+1)t}
\right]. \label{energy1}
\end{equation}
The associated ergotropy is then obtained as
\begin{equation}
\mathcal{W}_1(t)=0.
\end{equation}
Thus, the first important observation is that the squeezed-field charger
can transfer a finite amount of energy to the single-emitter battery
without rendering that energy extractable as work. The origin of this behavior follows from the structure of the
single-emitter state [Eq.~\eqref{DM1}]. As shown explicitly in
Appendix~\ref{app:single_emitter}, an emitter initially prepared in the ground state remains diagonal in the energy basis throughout the
evolution, while its excited-state population satisfies
$\rho_{ee}(t)<1/2$ for any finite squeezing strength. Consequently,
$\rho_{gg}(t)>\rho_{ee}(t)$, so the populations are already ordered
with the larger eigenvalue occupying the lower-energy state. The
density matrix is therefore passive with respect to $H_B$, and no
cyclic unitary transformation can further lower its energy. Hence, energy accumulation in the single-emitter battery does not imply the storage of unitary-extractable work.

This result is consistent with the standard characterization of
passivity, according to which a state is passive when it is diagonal
in the energy eigenbasis and its populations are nonincreasing with
increasing energy, so that no unitary transformation can further lower its average energy~\cite{campaioli2024colloquium}. For a two-level QB, the same conclusion is also consistent with
the general qubit ergotropy expression derived in
Ref.~\cite{yang2023battery}. For an incoherent state with excited-state
population $\rho_{ee}<1/2$, the state eigenvalues are already arranged
in passive order, and the ergotropy therefore vanishes.

The distinction becomes particularly transparent in the strong-squeezing
limit. For a fixed squeezing strength $r$, the long-time excited-state
population is
\begin{equation}
\rho_{ee}^{ss}
=
\frac{N}{2N+1}.
\end{equation}
In the strong-squeezing regime, $r\gg 1$, the mean photon number
$N=\sinh^2 r$ becomes large and the steady-state excited-state
population approaches $1/2$. The single-emitter battery therefore
stores approximately $\hbar\omega_0/2$, while its steady state becomes
nearly maximally mixed, $\rho_1^{\rm ss}\simeq \mathbb{I}_2/2$, which is passive. Thus, increasing the
squeezing strength raises the steady-state stored energy asymptotically
to one half of the maximum excitation energy of the single-emitter
battery, while the ergotropy remains zero. This illustrates the distinction between the internal energy stored in
the battery and the portion of that energy that is extractable through
unitary operations
\cite{campaioli2024colloquium,yang2023battery}.
The vanishing ergotropy found here should not be interpreted as a
generic limitation of single-qubit batteries. Indeed, engineered
dissipative processes can drive a single-qubit QB toward an
active equilibrium state with population inversion and finite
ergotropy \cite{barra2019dissipative}. In contrast, the present result
is specific to the squeezed-waveguide dynamics and the ground-state
initial condition considered here, which together lead to the passive
single-emitter evolution.

The single-emitter result therefore provides a natural reference point
for the collective configurations considered below. In a related
open-system setting, \c{C}akmak showed that collective coupling of
multiple qubits to a thermal environment can generate steady-state
coherences in the energy eigenbasis and finite ergotropy even in the
absence of population inversion
\cite{PhysRevE.102.042111}. Although the reservoir and coupling
mechanism considered there differ from the squeezed-waveguide model
studied here, this result illustrates how collective coupling can
qualitatively modify the work-storage properties of the global state.
In the present system, increasing the QB from one to two emitters
allows the phase-sensitive squeezed-reservoir terms to generate a
two-photon coherence between the collective states $|gg\rangle$ and
$|ee\rangle$. As shown in the following subsection, for suitable
emitter geometries this coherence renders the two-emitter state
nonpassive and gives rise to finite ergotropy.

\subsection{Two-emitter battery}
\label{sec:two}

We next consider the two-emitter QBs ($L=2$), initially
prepared in the collective ground state
$\rho_2(0)=|gg\rangle\langle gg|$. To obtain an analytically tractable
solution and identify the mechanism responsible for the emergence of
extractable work, we first focus on the configuration
$r_c=0$ and $r_{12}=\lambda_{0z}/4$. For this geometry,
$\gamma_{+}=\gamma_{-}=\gamma$ and
$\gamma_{12}^{\prime}=\gamma$, with $\gamma=\gamma_{1d}$.

In the collective basis
$\{|gg\rangle,|+\rangle,|-\rangle,|ee\rangle\}$, with
$|\pm\rangle=(|eg\rangle\pm|ge\rangle)/\sqrt{2}$, the dynamics of the density matrix
preserves the form
\begin{equation}
\rho_{2}(t)
=
\begin{pmatrix}
g(t) & 0 & 0 & \dfrac{u(t)}{2}e^{-i\phi}\\
0 & P(t) & 0 & 0\\
0 & 0 & P(t) & 0\\
\dfrac{u(t)}{2}e^{i\phi} & 0 & 0 & e(t)
\end{pmatrix},
\label{DM2}
\end{equation}
where $\phi=2k_{0z}R$. Here,
$d(t)= g(t)-e(t)$ describes the population imbalance between
the collective ground and doubly excited states, while $u(t)$ is the
phase-rotated two-photon coherence between $|gg\rangle$ and
$|ee\rangle$. The closed-form expressions for
$g(t)$, $e(t)$, $P(t)$, $d(t)$, and $u(t)$ are given in
Appendix~\ref{app:two_emitter}.

The stored energy takes the simple form
\begin{equation}
E_2(t)
=
\frac{2N}{2N+1}
\left[
1-e^{-(2N+1)\gamma t}
\right]
\hbar\omega_0. \label{2trane}
\end{equation}
Diagonalization of the density matrix given by Eq.~\eqref{DM2} and construction of its passive
counterpart, as detailed in Appendix~\ref{app:two_emitter}, give
\begin{equation}
\mathcal{W}_2(t)
=
\hbar\omega_0
\left[
\sqrt{d^2(t)+u^2(t)}
-d(t)
\right]. \label{2tranw}
\end{equation}

A direct comparison with the single-emitter result reveals an
important distinction between energy storage and work storage.
From Eq.~\eqref{energy1},
\begin{equation}
E_2(t)=2E_1(t),
\end{equation}
so that the energy deposited per emitter is identical in the
one- and two-emitter configurations considered here. The appearance
of finite ergotropy for two-emitter case therefore does not originate from an
enhancement of the absorbed energy per emitter. Instead, it results
from the different structure of the collective state. For $L=1$,
the density matrix [Eq.~\eqref{DM1}] remains diagonal with
$\rho_{gg}(t)>\rho_{ee}(t)$ and is therefore passive, giving
$\mathcal{W}_1(t)=0$. By contrast, for $L=2$ the phase-sensitive
squeezed-reservoir terms generate the coherence $u(t)$ between
$|gg\rangle$ and $|ee\rangle$. Since $d(t)>0$, the expression above
shows explicitly that any nonzero $u(t)$ increases
$\sqrt{d^2+u^2}$ above $d$ and therefore renders the state
nonpassive.
\begin{figure}[t]
    \centering  \includegraphics[width=0.49\columnwidth]{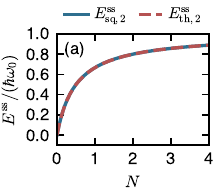}
    \hfill
\includegraphics[width=0.49\columnwidth]{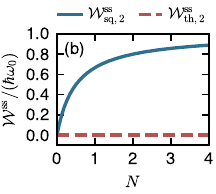}
    \caption{Comparison of the steady-state performance of the two-emitter QB coupled to squeezed-vacuum and thermal reservoirs with the same mean photon number $N$. (a) Steady-state stored energy, $E_{\rm sq}^{ss}$ and $E_{\rm th}^{ss}$. (b) Steady-state ergotropy, $\mathcal{W}_{\rm sq}^{ss}$ and $\mathcal{W}_{\rm th}^{ss}$. The two-emitter configuration is $r_c=0$ and $r_{12}=\lambda_{0z}/4$. }
    \label{fig1}
\end{figure}

The physical role of this coherence becomes particularly transparent
in the steady state. In the steady-state regime, the single-excitation population vanishes,
$P^{\rm ss}=0$, while the remaining populations and two-photon
coherence are given by $g^{\rm ss}=\frac{N+1}{2N+1},$ $e^{\rm ss}=\frac{N}{2N+1}$ and $u^{\rm ss}
=
-\frac{2\sqrt{N(N+1)}}{2N+1}.$

The corresponding density matrix is pure and can be written as $|\Psi_{\rm sq}^{\rm ss}\rangle
=
\frac{
\sqrt{N+1}\,|gg\rangle
-
e^{i\phi}\sqrt{N}\,|ee\rangle
}{
\sqrt{2N+1}
}.$ This form is consistent with the  two-emitter steady state solution
obtained for squeezed-vacuum waveguide QED, where suitable
center-of-mass positions suppress the populations of the
single-excitation states and stabilize a coherent superposition of
$|gg\rangle$ and $|ee\rangle$~\cite{you2018waveguide}. In the
present configuration, the spectrum of the steady state is
$\{1,0,0,0\}$. Since the battery Hamiltonian has a unique collective
ground state, the passive rearrangement places the single nonzero
eigenvalue on $|gg\rangle$, giving zero passive energy. It follows
that
\begin{equation}
E_{\rm sq,2}^{\rm ss}
=
\mathcal{W}_{\rm sq,2}^{\rm ss}
=
\frac{2N}{2N+1}\hbar\omega_0. \label{ssenew}
\end{equation}
Thus, for this geometry, the energy accumulated in the steady state
is entirely unitarily extractable.

This result naturally raises the question of whether the finite
ergotropy originates simply from the nonzero mean occupation $N$ of
the reservoir or from the phase-sensitive correlations specific to
the squeezed vacuum. The distinction can be examined directly by
comparing with a thermal reservoir having the same $N$. In the
master equation, the ordinary absorption and emission contributions
have the same structure as those of a thermal reservoir, whereas
the squeezed vacuum introduces the additional phase-sensitive
two-photon contribution proportional to
$M$~\cite{you2018waveguide}. Therefore, we compare the squeezed reservoir with a thermal reservoir
having the same mean photon number $N$. Within Eq.~\eqref{eqmaster}, the normal
absorption and emission contributions proportional to $N$ and $N+1$
are then identical for the two reservoirs, whereas the anomalous
phase-sensitive correlations proportional to $M$ are present only for
the squeezed vacuum.

For the present separation $r_{12}=\lambda_{0z}/4$, one also has
$\gamma_{12}=0$, and the thermal steady state takes the compact
product form $\rho_{\rm th,2}^{\rm ss}
=
\left[
\frac{
(N+1)|g\rangle\langle g|
+
N|e\rangle\langle e|
}{
2N+1
}
\right]^{\otimes 2}$~\cite{you2018waveguide}. Its stored energy is
\begin{equation}
E_{\rm th,2}^{\rm ss}
=
\frac{2N}{2N+1}\hbar\omega_0,
\end{equation}
which is identical to the squeezed-reservoir result. However,
$\rho_{\rm th,2}^{\rm ss}$ is diagonal in the energy basis and its
populations decrease with increasing excitation energy. It is
therefore already passive, and
\begin{equation}
\mathcal{W}_{\rm th,2}^{\rm ss}=0.
\end{equation}

Figure~\ref{fig1} makes the distinction between stored energy and
extractable work explicit. At fixed mean photon number $N$, the
squeezed and thermal reservoirs produce the same steady-state energy
[Fig.~\ref{fig1}(a)], but markedly different ergotropies. The thermal
steady state remains passive, whereas the squeezed reservoir stabilizes
a coherent $|gg\rangle$--$|ee\rangle$ superposition for which
$\mathcal{W}_{\rm sq,2}^{\rm ss}=E_{\rm sq,2}^{\rm ss}$
[Fig.~\ref{fig1}(b)]. The difference therefore does not originate from
a larger amount of deposited energy, but from the phase-sensitive
two-photon correlations of the squeezed reservoir, which reorganize
the steady state and render the stored energy unitarily
extractable.

This interpretation is consistent with the broader relation between
coherence and extractable work. Francica \textit{et al.} showed that
ergotropy can be separated into incoherent and coherent contributions,
with the latter associated with coherence in the energy eigenbasis
~\cite{francica2020quantum}. In an open-system setting, \c{C}akmak
showed that collective coupling to a thermal environment can generate
steady-state coherence and finite ergotropy even in the absence of
population inversion~\cite{PhysRevE.102.042111}. These results
emphasize that nonpassivity is determined not only by the energy
supplied by the environment, but also by the structure of the state
generated by the system--reservoir interaction. %In the present case, the phase-sensitive $M$-dependent processes provide precisely this additional mechanism by sustaining coherence between different energy sectors.

These results establish the two-emitter system as the minimal
collective configuration in which phase-sensitive reservoir
correlations can be converted into extractable work. We next consider
three emitters to examine how this mechanism extends to a larger
collective Hilbert space.

\subsection{Three-emitter battery}
\label{sec:three_emitter}

Now, We extend the analysis to a three-emitter battery in order to
determine how the collective charging mechanism develops beyond the
minimal two-emitter configuration. The emitters are placed
symmetrically with respect to the center of the squeezing sources, $r_1=-\lambda_{0z},$ $r_2=0$ and $r_3=\lambda_{0z},$ where $\lambda_{0z}=2\pi/k_{0z}$ is the effective longitudinal
wavelength of the waveguide. For this configuration, $\gamma_{ij}=\gamma'_{ij}=\gamma$ and $\Omega_{ij}=0,$ so that both the ordinary dissipative coupling and the
phase-sensitive squeezed-reservoir coupling are fully collective.

For this symmetric configuration, the master equation is invariant
under exchange of the emitter labels. Together with the fully
symmetric initial state $\ket{ggg}$, this ensures that the dynamics
remains confined to the permutation-symmetric subspace. We introduce the collective basis $|0\rangle=|ggg\rangle,$ $|1\rangle=\frac{|egg\rangle+|geg\rangle+|gge\rangle}{\sqrt{3}},$ $|2\rangle=\frac{|eeg\rangle+|ege\rangle+|gee\rangle}{\sqrt{3}}$ and $|3\rangle=|eee\rangle.$

As shown in Appendix~\ref{app:three_emitter}, excitation-parity
symmetry further restricts the density matrix to
\begin{equation}
\rho_3(t)=
\begin{pmatrix}
P_0(t) & 0 & e^{-i\phi}X(t) & 0\\
0 & P_1(t) & 0 & e^{-i\phi}Y(t)\\
e^{i\phi}X(t) & 0 & P_2(t) & 0\\
0 & e^{i\phi}Y(t) & 0 & P_3(t)
\end{pmatrix},
\label{eq:rho3_main}
\end{equation}
where $P_n$ denotes the population of the $n$-excitation symmetric
state and $X=e^{i\phi}\rho_{02}$ and $Y=e^{i\phi}\rho_{13},$ are real phase-rotated coherences. %The complete six-dimensional system of equations governing $\{P_0,P_1,P_2,P_3,X,Y\}$ is given in Appendix~\ref{app:three_emitter}. 
The stored energy follows directly from the excitation populations,
\begin{equation}
E_3(t)
=
\hbar\omega_0
\left[
P_1(t)+2P_2(t)+3P_3(t)
\right].
\label{3trane}
\end{equation}
In contrast, evaluation of the ergotropy requires the spectrum of the
density operator over the complete eight-dimensional Hilbert space of
the three-emitter battery. The four nonzero eigenvalues arise from
the two independent blocks associated with the even- and odd-parity
sectors,
\begin{subequations}
\begin{align}
\lambda_{02,\pm}
&=
\frac{
P_0+P_2
\pm
\sqrt{(P_0-P_2)^2+4X^2}
}{2},
\\
\lambda_{13,\pm}
&=
\frac{
P_1+P_3
\pm
\sqrt{(P_1-P_3)^2+4Y^2}
}{2}.
\end{align}
\label{eq:rho3_eigenvalues_main}
\end{subequations}
The remaining four eigenvalues are zero because the evolution has no
support outside the symmetric subspace.

\begin{figure}[t]
    \centering  \includegraphics[width=0.49\columnwidth]{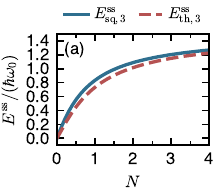}
    \hfill
\includegraphics[width=0.49\columnwidth]{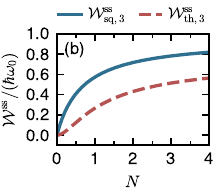}
   \caption{Comparison of the steady-state performance of the three-emitter QB coupled to squeezed-vacuum and thermal reservoirs with the same mean photon number $N$. (a) Steady-state stored energy, $E_{\rm sq}^{ss}$ and $E_{\rm th}^{ss}$. (b) Steady-state ergotropy, $\mathcal{W}_{\rm sq}^{ss}$ and $\mathcal{W}_{\rm th}^{ss}$. The three-emitter configuration is $r_1=-\lambda_{0z}$, $r_2=0$, and $r_3=\lambda_{0z}$.}
    \label{fig2}
\end{figure}

Let $\nu_1(t)
=
\max
\left\{
\lambda_{02,+}(t),
\lambda_{13,+}(t)
\right\}$ denote the largest eigenvalue of $\rho_3(t)$. Since the physical
three-emitter Hamiltonian contains a unique ground state followed by
a threefold-degenerate one-excitation manifold, the passive energy is $E_{\mathrm{pas},3}(t)
=
\hbar\omega_0
\left[
1-\nu_1(t)
\right].$

The three-emitter ergotropy therefore takes the compact form
\begin{equation}
\mathcal{W}_3(t)
=
\hbar\omega_0
\left[
P_1(t)+2P_2(t)+3P_3(t)-1+\nu_1(t)
\right].
\label{3tranw}
\end{equation}

Equation~\eqref{eq:rho3_main} reveals an important difference from
the two-emitter case. The three-emitter dynamics supports two
phase-sensitive coherence channels,
$|0\rangle\leftrightarrow|2\rangle$ and
$|1\rangle\leftrightarrow|3\rangle$, represented by $X$ and $Y$,
respectively. These coherences enter the eigenvalue spectrum of the
battery state and therefore affect its passive-energy contribution and
ergotropy.

We now turn to the steady-state regime and examine how the
phase-sensitive correlations of the squeezed reservoir affect the
battery performance. To distinguish this contribution from that of
ordinary collective thermalization, we compare the squeezed reservoir
with a thermal reservoir having the same mean occupation number $N$.
%In the thermal case, the anomalous correlations proportional to $M$ are absent, while the absorption and emission contributions proportional to $N$ and $N+1$ remain~\cite{you2018waveguide}.
For the three-emitter geometry considered here, all pairwise
separations are integer multiples of the resonant wavelength, yielding
$\gamma_{ij}=\gamma$ and $\Omega_{ij}=0$. Starting from the same initial condition as squeezed reservoir, the corresponding thermal master equation
does not generate coherences between the symmetric Dicke states, and
the steady state is therefore diagonal in the basis
$\{|0\rangle,|1\rangle,|2\rangle,|3\rangle\}$. Solving the population equations in the steady-state limit gives $
P_0^{\rm th}
=
\frac{(N+1)^3}
{4N^3+6N^2+4N+1},
$ $P_1^{\rm th}
=
\frac{N(N+1)^2}
{4N^3+6N^2+4N+1},
$ $
P_2^{\rm th}
=
\frac{N^2(N+1)}
{4N^3+6N^2+4N+1},
$ and $
P_3^{\rm th}
=
\frac{N^3}
{4N^3+6N^2+4N+1}.
$

Although these populations have been obtained here directly from the
thermal limit of the waveguide master equation, they coincide with the
steady-state population structure found for collectively coupled
emitters interacting with a thermal reservoir
~\cite{PhysRevE.102.042111,ahmadi2025harnessing}. The physical
realization of the collective regime is, however, different. Its steady-state energy follows directly from the excitation
populations,
\begin{equation}
E_{\rm th,3}^{\rm ss}
=
\frac{
N(6N^2+4N+1)
\hbar\omega_0}{
(2N+1)(2N^2+2N+1)
}.
\label{eq:E_th_L3}
\end{equation}
Although the thermal steady state is diagonal in the symmetric Dicke
basis, it is not passive with respect to the full three-emitter
Hamiltonian. Its spectrum is
$\{P_0^{\rm th},P_1^{\rm th},P_2^{\rm th},P_3^{\rm th},0,0,0,0\}$,
whereas the full Hamiltonian has the energy spectrum
$\{0,\hbar\omega_0,\hbar\omega_0,\hbar\omega_0,
2\hbar\omega_0,2\hbar\omega_0,2\hbar\omega_0,
3\hbar\omega_0\}$. The corresponding
passive energy is therefore
$E_{\rm pas,th,3}^{\rm ss}
=\hbar\omega_0(P_1^{\rm th}+P_2^{\rm th}+P_3^{\rm th})$.
Consequently, the steady-state ergotropy, yields
\begin{equation}
\mathcal{W}_{\rm th,3}^{\rm ss}
=
\frac{
N^2(3N+1)\hbar\omega_0
}{
(2N+1)(2N^2+2N+1)
}.
\label{eq:W_th_L3}
\end{equation}
Figure~\ref{fig2} compares the steady-state performance of the
three-emitter battery under thermal and squeezed reservoirs with the
same mean occupation number $N$. We first consider the stored energy,
shown in Fig.~\ref{fig2}(a). Both reservoirs charge the battery to a
finite steady-state energy, with the squeezed reservoir yielding a
higher value over the range considered. The difference, however,
remains comparatively modest. We next consider the ergotropy, shown in Fig.~\ref{fig2}(b). Starting
from zero at $N=0$, the ergotropy increases with the reservoir
occupation for both the thermal and squeezed cases. In contrast to the
two-emitter configuration, the thermal curve is already finite for
$N>0$, reflecting the nonpassive steady state generated by the
collective three-emitter dynamics. The squeezed reservoir produces a
systematically larger ergotropy over the range considered, and the
separation between the two curves is substantially more pronounced
than for the stored energy in Fig.~\ref{fig2}(a). Thus, this indicates that the principal effect of the
phase-sensitive two-photon correlations proportional to $M$ is not simply to
increase the total energy stored in the battery, but to modify the
steady-state spectrum in a way that reduces its passive contribution
and increases the fraction of stored energy that remains unitarily
extractable.

\section{Control of collective QB performance}
\label{sec4}

\subsection{Transient charging dynamics}

\begin{figure}[!htbp]
    \centering \includegraphics[width=\columnwidth]{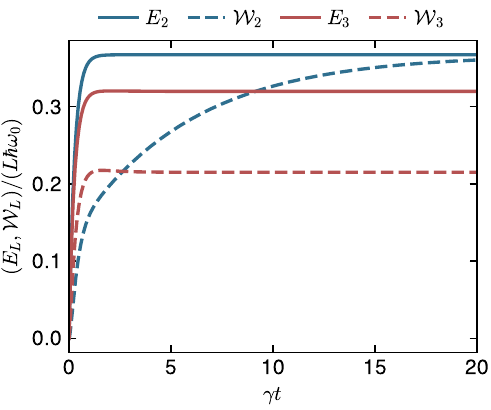}
\caption{
Transient charging dynamics of the collective QBs for
$L=2$ and $L=3$ emitters at squeezing parameter $r=1$.
The stored energy $E_L/(L\hbar\omega_0)$ (solid lines) and ergotropy
$\mathcal{W}_L/(L\hbar\omega_0)$ (dashed lines) are shown as functions
of the dimensionless time $\gamma t$. The two-emitter configuration is
$r_c=0$ and $r_{12}=\lambda_{0z}/4$, while for three emitters
$r_1=-\lambda_{0z}$, $r_2=0$, and $r_3=\lambda_{0z}$.
}\label{fig3}
\end{figure}

The steady-state analysis characterizes the final energetic
performance of the batteries configuration, but does not reveal the time scales over
which stored energy and extractable work are established. We therefore
examine the transient charging dynamics of the two- and three-emitter
configurations introduced in Secs.~\ref{sec:two} and~\ref{sec:three_emitter}. Figure~\ref{fig3}
shows the stored energy $E_L/(L\hbar\omega_0)$ and ergotropy
$\mathcal{W}_L/(L\hbar\omega_0)$ for a squeezing parameter $r=1$.

For two emitters, Eq.~\eqref{2trane} shows that the stored energy increases
monotonically according to a single exponential with decay rate
$(2N+1)\gamma$. Consequently, the energy approaches its steady-state
value on the time scale
$[(2N+1)\gamma]^{-1}$ and is already close to saturation for
$\gamma t\sim 1$. The ergotropy, however, exhibits a markedly slower
approach to its asymptotic value. According to Eq.~\eqref{2tranw},
$\mathcal{W}_2(t)$ depends not only on the population imbalance
$d(t)$ but also on the two-photon coherence $u(t)$. Thus, although
energy is transferred rapidly from the reservoir to the emitters, its
conversion into fully extractable work requires the slower
reorganization of the collective state. In particular, the transient
single-excitation population must decay while the
$|gg\rangle\leftrightarrow|ee\rangle$ coherence approaches its
steady state value. This accounts for the continued growth of
$\mathcal{W}_2$ after $E_2$ has essentially reached its plateau.

The three-emitter dynamics displays a different separation of time
scales. From Eqs.~\eqref{3trane} and~\eqref{3tranw}, $E_3(t)$ is determined solely by the
excitation populations, whereas $\mathcal{W}_3(t)$ also depends on the
largest eigenvalue $\nu_1(t)$ of the density operator and therefore on
the two coherence channels $X(t)$ and $Y(t)$. As shown in
Fig.~\ref{fig3}, both $E_3$ and $\mathcal{W}_3$ rise rapidly and reach
values close to their steady state limits on comparable time scales. The three-emitter ergotropy briefly exceeds its steady-state value
before relaxing, reflecting the coupled evolution of populations and
phase-sensitive coherences in the two parity sectors.

A useful comparison follows from the short-time behavior of the
stored energy. For both configurations,
\begin{equation}
\left.
\frac{d}{d(\gamma t)}
\frac{E_L(t)}{L\hbar\omega_0}
\right|_{t=0}
=
N,
\qquad L=2,3.
\label{eq:initial_charging_rate}
\end{equation}
Thus, both QBs exhibit the same initial charging rate per emitter. Their subsequent dynamics differ because collective
population redistribution and coherence generation proceed
differently as the size of the symmetric excitation manifold
increases.

At $r=1$, the two-emitter battery reaches
$E_2^{\rm ss}=\mathcal{W}_2^{\rm ss}$, whereas the three-emitter
battery retains a finite passive contribution,
$\mathcal{W}_3^{\rm ss}<E_3^{\rm ss}$. Thus, increasing the number of
emitters changes not only the transient generation of nonpassivity but
also the fraction of stored energy that remains extractable. Since the
collective couplings in the waveguide depend explicitly on the emitter
positions, the next natural step is to examine how geometry can be
used to control this balance between stored energy and ergotropy.

\subsection{Geometric and reservoir control}

\begin{figure}[!htbp]
     \centering  \includegraphics[width=0.49\columnwidth]{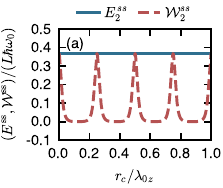}
    \hfill
\includegraphics[width=0.49\columnwidth]{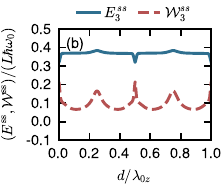}
\caption{Steady-state energy and ergotropy for two- and three-emitter QBs at squeezing parameter $r=1$. (a) Two-emitter steady-state energy $E_2^{ss}$ and ergotropy $\mathcal{W}_2^{ss}$ as functions of the normalized center-of-mass position $r_c/\lambda_{0z}$, with fixed separation $r_{12}=\lambda_{0z}/4$. (b) Three-emitter steady-state energy $E_3^{ss}$ and ergotropy $\mathcal{W}_3^{ss}$ as functions of the normalized spacing $d/\lambda_{0z}$ for the symmetric configuration $r_1=-d$, $r_2=0$, and $r_3=d$. Energies are given in units of $\hbar\omega_0$.}
    \label{fig4}
\end{figure}

In this subsection, we exploit the explicit spatial dependence of the waveguide
couplings to determine how the energetic performance of the battery
can be controlled by geometry alone. According to Eqs.~\eqref{gammaij}--\eqref{gammapij},
the ordinary collective decay and coherent exchange depend on relative
emitter separations, whereas the phase-sensitive squeezed-reservoir
coupling depends on the sums of the emitter positions. Geometry
therefore provides a means of modifying the phase-sensitive
correlations independently of, or together with, the ordinary
collective coupling. To study this effect from changes in reservoir
strength, the squeezing parameter is fixed at $r=1$ throughout
Fig.~\ref{fig4}.

For two emitters, we keep the separation fixed at
$r_{12}=\lambda_{0z}/4$ and translate the pair by varying its
center-of-mass position $r_c$. The relative-position-dependent
coefficients are consequently unchanged, with
$\gamma_{12}=0$, while the phase-sensitive coefficients entering the
two-emitter dynamics vary as
$\gamma'_{12}=\gamma'_+=\gamma'_-
=\gamma\cos(2k_{0z}r_c)$. Importantly, the squeezed contribution
cancels from the equation governing the population imbalance
$d=g-e$, so that its steady-state value, and hence the stored energy,
is independent of $r_c$. This accounts for the constant
$E_2^{\rm ss}$ in Fig.~\ref{fig4}(a).

The ergotropy displays a qualitatively different behavior. At
$r_c=m\lambda_{0z}/4$, with integer $m$, the magnitude of the
phase-sensitive coupling is maximal,
$|\gamma'_{12}|=\gamma$, and the steady state differs from that
obtained at $r_c=0$ only through the sign or phase of its two-photon
coherence. Its spectrum is therefore unchanged, and the battery
reaches $\mathcal{W}_2^{\rm ss}=E_2^{\rm ss}$. By contrast, at
$r_c=(2m+1)\lambda_{0z}/8$, the phase-sensitive coupling vanishes.
The $|gg\rangle\leftrightarrow|ee\rangle$ coherence is then absent,
and the resulting steady state is passive, giving
$\mathcal{W}_2^{\rm ss}=0$. Thus, translating the emitter pair can
continuously switch the steady-state ergotropy between zero and the
entire stored energy without changing the amount of energy deposited
in the battery. This provides a direct separation between geometric
control of energy storage and geometric control of energy
extractability.

The three-emitter configuration provides a more intricate case
because varying the symmetric spacing,
$r_1=-d$, $r_2=0$, and $r_3=d$, modifies both the ordinary and
phase-sensitive couplings. In particular,
$\gamma_{12}=\gamma_{23}=\gamma\cos(k_{0z}d)$ and
$\gamma_{13}=\gamma\cos(2k_{0z}d)$, while
$\Omega_{12}=\Omega_{23}=(\gamma/2)\sin(k_{0z}d)$ and
$\Omega_{13}=(\gamma/2)\sin(2k_{0z}d)$. At the same time,
$\gamma'_{12}=\gamma'_{23}=\gamma\cos(k_{0z}d)$,
$\gamma'_{13}=\gamma$, and the diagonal phase-sensitive coefficients
of the outer emitters vary as $\gamma\cos(2k_{0z}d)$. The spacing
therefore changes several competing channels simultaneously, in
contrast to the rigid translation considered for $L=2$.

This distinction is reflected in Fig.~\ref{fig4}(b). The stored
energy remains comparatively weakly dependent on $d$, whereas the
ergotropy exhibits a substantially stronger spatial modulation.
Particularly pronounced features occur at
$d=m\lambda_{0z}/2$, where the coherent exchange terms $\Omega_{ij}$ vanish and
the dissipative couplings $\gamma_{ij}$ reach their extremal magnitudes. The
transformation $d\rightarrow d+\lambda_{0z}/2$ changes the sign of
the couplings involving the central emitter but is equivalent to a
local $\pi$ phase rotation of that emitter. Since this transformation
leaves the bare battery Hamiltonian invariant, the energetic
quantities inherit the corresponding half-wavelength periodicity.
Secondary ergotropy maxima near quarter-wavelength spacings occur
where the nearest-neighbor dissipative couplings vanish while the
corresponding coherent exchange is maximal, illustrating that the
three-emitter response results from the interplay of several
geometry-dependent channels rather than from a single collective
rate.

Figure~\ref{fig4} therefore demonstrates that the waveguide geometry
acts primarily as a control parameter for the structure and
nonpassivity of the steady state. In particular, substantial
changes in ergotropy can occur with comparatively small changes in
stored energy. Geometry can thus redistribute the deposited energy
between passive and unitarily extractable components without requiring
a change in the reservoir strength. 

\begin{figure}[!htbp]
      \centering  \includegraphics[width=0.49\columnwidth]{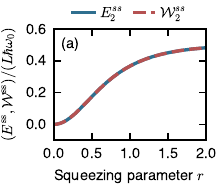}
    \hfill
\includegraphics[width=0.49\columnwidth]{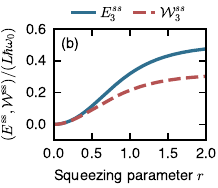}
\caption{Steady-state energy and ergotropy as functions of the squeezing parameter $r$. (a) Two-emitter case: $E_2^{ss}$ and $\mathcal{W}_2^{ss}$ for the configuration $r_c=0$ and $r_{12}=\lambda_{0z}/4$. In this case, the steady-state ergotropy coincides with the stored energy, i.e., $\mathcal{W}_2^{ss}=E_2^{ss}$. (b) Three-emitter case: $E_3^{ss}$ and $\mathcal{W}_3^{ss}$ for the symmetric configuration $r_1=-\lambda_{0z}$, $r_2=0$, and $r_3=\lambda_{0z}$. Energies are given in units of $\hbar\omega_0$.}
    \label{fig5}
\end{figure}

Once we established this spatial
control, we next examine how the same energetic quantities evolve when
the squeezing strength itself is varied. Figure~\ref{fig5} completes this analysis by varying the squeezing
parameter while keeping the representative geometries fixed. For both
batteries configuration, the stored energy and ergotropy vanish at $r=0$ and increase
monotonically as $N=\sinh^2 r$ and
$M=\sinh r\cosh r$ grow. For $L=2$, the relation
$E_2^{\rm ss}=\mathcal{W}_2^{\rm ss}$ remains valid for all $r$, as
follows from Eq.~\eqref{ssenew}, showing that increasing the squeezing strength
changes the amount of stored energy without introducing a passive
contribution. For $L=3$, both quantities also increase with $r$, but
$\mathcal{W}_3^{\rm ss}$ remains below $E_3^{\rm ss}$ throughout the
range considered, indicating that a finite fraction of the stored
energy remains passive even as the phase-sensitive correlations become
stronger. Thus, geometry and squeezing play complementary roles:
geometry determines how efficiently reservoir correlations are
converted into nonpassivity, whereas the squeezing strength controls
the overall scale of the charging process.

\section{Coherence and entanglement}
\label{sec5}

Now we consider how nonclassical resources contribute to the behavior
described above. We characterize the quantum properties generated by
the squeezed reservoir by considering the $l_1$-norm coherence and the
pairwise concurrence. The $l_1$-norm coherence is defined as~\cite{PhysRevLett.113.140401,streltsov2017colloquium}
\begin{equation}
C_{l_1}(\rho_S)
=
\sum_{m\neq n}
\left|
(\rho_S)_{mn}
\right|.
\label{eq:l1}
\end{equation}

Throughout this work, $C_{l_1}$ is evaluated in the computational
product basis $\{|g\rangle,|e\rangle\}^{\otimes L}$, which is also an
eigenbasis of the bare battery Hamiltonian $H_B$. Bipartite entanglement is quantified by the concurrence. For a pair
of emitters $(i,j)$,
\begin{equation}
\mathcal{C}_{ij}
=
\max\left\{
0,\lambda_1-\lambda_2-\lambda_3-\lambda_4
\right\},
\end{equation}
where the $\lambda_i$, in decreasing order, are the eigenvalues of
$\mathcal{R}
=\sqrt{\sqrt{\rho_{ij}}\,\widetilde{\rho}_{ij}\sqrt{\rho_{ij}}}$,
with
$\widetilde{\rho}_{ij}
=(\sigma_y\otimes\sigma_y)\rho_{ij}^{*}
(\sigma_y\otimes\sigma_y)$.
For $L=2$, $\rho_{ij}$ is the full two-emitter state, whereas for
$L=3$ it is obtained by tracing over the remaining emitter,
$\rho_{ij}=\operatorname{Tr}_{k}\rho_3$.

For the two-emitter state in Eq.~\eqref{DM2}, the $X$ structure of the
density matrix reduces the concurrence to
\begin{equation}
\mathcal{C}_{12}(t)
=
\max\left\{0,|u(t)|-2P(t)\right\}.
\label{eq:C12_L2}
\end{equation}
At the collective steady state considered is given by,
$P^{\rm ss}=0$, and therefore
$\mathcal{C}_{12}^{\rm ss}=|u^{\rm ss}|$.

For three emitters, the concurrence is evaluated from the corresponding
two-emitter reduced density matrices. For the symmetric arrangement
$r_1=-d$, $r_2=0$, and $r_3=d$, reflection symmetry implies
$\mathcal{C}_{12}=\mathcal{C}_{23}$, while
$\mathcal{C}_{13}$ independently characterizes the correlations
between the two outer emitters.

\begin{figure}[!htbp]
      \centering  \includegraphics[width=0.49\columnwidth]{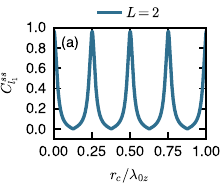}
    \hfill
\includegraphics[width=0.49\columnwidth]{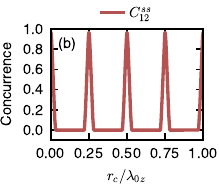}
\includegraphics[width=0.49\columnwidth]{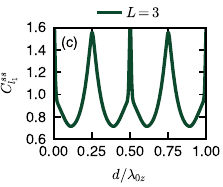}
    \hfill
\includegraphics[width=0.49\columnwidth]{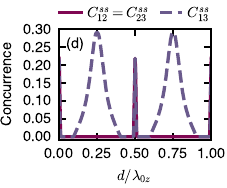}
\caption{Steady-state coherence and pairwise entanglement for two- and three-emitter QBs at squeezing parameter $r=1$. (a) $l_1$-norm coherence $C_{l_1}^{ss}$ and (b) concurrence $C_{12}^{ss}$ for the two-emitter system as functions of the normalized center-of-mass position $r_c/\lambda_{0z}$, with fixed separation $r_{12}=\lambda_{0z}/4$. (c) $l_1$-norm coherence $C_{l_1}^{ss}$ and (d) pairwise concurrences for the three-emitter system as functions of the normalized spacing $d/\lambda_{0z}$ for the symmetric configuration $r_1=-d$, $r_2=0$, and $r_3=d$. Owing to reflection symmetry, $C_{12}^{ss}=C_{23}^{ss}$, while $C_{13}^{ss}$ describes the concurrence between the two outer emitters.}
    \label{fig6}
\end{figure}

Figure~\ref{fig6} allows us to examine whether the geometric
enhancement of ergotropy identified in Fig.~\ref{fig4} is accompanied
by corresponding changes in coherence and entanglement. For the
two-emitter battery, the connection is particularly direct. Since the
only off-diagonal contribution in the computational energy basis is
the $|gg\rangle\leftrightarrow|ee\rangle$ coherence, one has
$C_{l_1}=|u|$, while Eq.~\eqref{eq:C12_L2} gives
$\mathcal{C}_{12}=\max\{0,|u|-2P\}$. As shown in
Figs.~\ref{fig6}(a) and \ref{fig6}(b), both quantities are strongly
modulated by the center-of-mass position and attain their maxima at
$r_c=m\lambda_{0z}/4$, precisely where the phase-sensitive coupling
has maximal magnitude. For $r_c=m\lambda_{0z}/4$, the steady-state single-excitation
population vanishes, $P^{\rm ss}=0$, so that
$\mathcal{C}_{12}^{\rm ss}=C_{l_1}^{\rm ss}=|u^{\rm ss}|$. These
points coincide with the maxima of the ergotropy in
Fig.~\ref{fig4}(a). Conversely, at
$r_c=(2m+1)\lambda_{0z}/8$, the anomalous coupling vanishes, the
two-photon coherence and concurrence disappear, and the ergotropy
simultaneously falls to zero, despite the stored energy remaining
unchanged. The two-emitter system therefore provides a clear example in which geometry controls extractable work
through the phase-sensitive coherence generated between different
energy sectors. This behavior is consistent with the general result
that coherence in the energy eigenbasis can provide a genuine
contribution to ergotropy~\cite{francica2020quantum}.

For the three-emitter battery, the steady-state $l_1$-norm coherence
shows a pronounced geometry dependence that closely follows the
ergotropy profile. As shown in Fig.~\ref{fig6}(c), coherence exhibits
maxima near $d/\lambda_{0z}=1/4$, $1/2$, and $3/4$, coinciding with the
ergotropy enhancements in Fig.~\ref{fig4}(b), with the dominant maximum
of both quantities occurring at $d/\lambda_{0z}=1/2$. This spatial
correspondence indicates that the geometries that most strongly enhance
the phase-sensitive coherences also favor the generation of
nonpassivity and, consequently, extractable work. At the same time, the correspondence
should not be interpreted as a universal proportionality, since the
$l_1$-norm is basis dependent and includes all off-diagonal elements
of the density matrix~\cite{PhysRevLett.113.140401,streltsov2017colloquium}. The concurrences provides a qualitatively different picture.
As shown in Fig.~\ref{fig6}(d), the dominant entangled pair changes
with the emitter spacing. Near $d=\lambda_{0z}/4$ and
$3\lambda_{0z}/4$, the outer-emitter concurrence
$\mathcal{C}_{13}$ is enhanced, while
$\mathcal{C}_{12}=\mathcal{C}_{23}$ is strongly suppressed. Around
$d=\lambda_{0z}/2$, where the ergotropy reaches its dominant maximum,
the pairwise concurrences do not exhibit a corresponding maximum.
Thus, although geometry redistributes pairwise entanglement among the
emitters, concurrence does not provide a direct or monotonic indicator
of extractable work. This behavior is consistent with the general
result that entanglement need not provide a positive contribution to
ergotropy and can even reduce the extractable work for mixed states
~\cite{PhysRevE.105.L052101}. 

In general, the comparison identifies coherence as the nonclassical
quantity most closely associated with the geometry-controlled
ergotropy in the present system, as evidenced by their coincident
spatial maxima. Pairwise entanglement, in contrast, does not track the
ergotropy monotonically and can vanish at geometries where finite
extractable work remains. 

%This distinction is consistent with the general result that entanglement does not necessarily provide a positive contribution to ergotropy and may even reduce the extractable work for mixed states~\cite{PhysRevE.105.L052101}.

\section{Conclusion}
\label{sec6}

We investigated the charging and work-storage properties of
few-emitter QBs coupled to a one-dimensional waveguide
driven by a squeezed-vacuum reservoir, with particular emphasis on the
role of collective coupling and emitter geometry. A single emitter stores a finite amount of energy but remains passive,
so no work can be extracted through cyclic unitary operations. For two
emitters, an appropriate geometry stabilizes a pure coherent steady
state for which the entire stored energy is extractable, despite a thermal reservoir
with the same mean occupation producing the same energy but zero
ergotropy. For three emitters, collective thermal dynamics already
produces finite ergotropy, while squeezing further enhances the
extractable component of the stored energy. The transient analysis
also showed that energy accumulation and the generation of
ergotropy can occur on distinct time scales. Most importantly, the waveguide geometry provides direct control over
the battery performance, allowing substantial modulation of ergotropy
with comparatively small changes in stored energy. In the two-emitter
case, geometry can switch the steady state between passive and fully
extractable regimes, while in the three-emitter case it produces a
strong spatial modulation of ergotropy through the interplay of
collective dissipative, coherent, and phase-sensitive couplings. The
coherence closely follows the spatial variation of the ergotropy in
both cases, whereas pairwise entanglement in the three-emitter battery
is redistributed among different emitter pairs and does not vary
monotonically with the extractable work.

These results identify emitter geometry and phase-sensitive reservoir
correlations as complementary resources for controlling the useful
energetic content of collective QBs. Extensions to
larger emitter arrays will be important for determining how the
geometry-controlled mechanism scales with system size and whether
collective advantages persist beyond the few-emitter regime. It would
also be relevant to examine the robustness of these effects against
imperfect positioning, additional loss channels, finite-bandwidth
squeezing, and other experimentally realistic imperfections. Such
studies may help establish waveguide reservoir engineering as a route
toward spatially programmable quantum-energy storage.

\appendix

\section{Single-emitter dynamics}
\label{app:single_emitter}

For completeness, we derive here the single-emitter solution used in
Sec.~\ref{sec:single}. Starting from the general master equation
\eqref{eqmaster} and setting $L=1$, the coherent dipole--dipole term
vanishes and the dynamics reduces to
\begin{align}
\frac{d\rho_1}{dt}
={}&
M\gamma'
\left(
e^{i\phi}S^{+}\rho_1 S^{+}
+
e^{-i\phi}S^{-}\rho_1 S^{-}
\right)
\nonumber\\
&-\frac{\gamma}{2}(N+1)
\left(
\rho_1 S^{+}S^{-}
+
S^{+}S^{-}\rho_1
-
2S^{-}\rho_1 S^{+}
\right)
\nonumber\\
&-\frac{\gamma}{2}N
\left(
\rho_1 S^{-}S^{+}
+
S^{-}S^{+}\rho_1
-
2S^{+}\rho_1 S^{-}
\right),
\label{eq:app_single_master}
\end{align}
where $\phi=2k_{0z}R,$ $\gamma=\gamma_{1d}$ and
$\gamma'=\gamma_{1d}\cos(2k_{0z}\delta)$ for an emitter located at
position $\delta$ along the waveguide.

In the basis $\{|e\rangle,|g\rangle\}$, we parametrize the
single-emitter density matrix as
\begin{equation}
\rho_{1}(t)
=
\begin{pmatrix}
p(t) & c(t)\\
c^{*}(t) & 1-p(t)
\end{pmatrix},
\label{eq:app_single_rho}
\end{equation}
where
$p(t)=\langle e|\rho_{1}(t)|e\rangle$
is the excited-state population and
$c(t)=\langle e|\rho_{1}(t)|g\rangle$
is the coherence between the excited and ground states. 
Taking the diagonal matrix element of
Eq.~\eqref{eq:app_single_master} gives
\begin{equation}
\dot{p}
=
\gamma N
-
\gamma(2N+1)p.
\label{eq:app_population2}
\end{equation}
For the initial condition $p(0)=0$, the solution is
\begin{equation}
p(t)
=
\frac{N}{2N+1}
\left[
1-e^{-\gamma(2N+1)t}
\right].
\label{eq:app_population_solution}
\end{equation}

The phase-sensitive squeezed-reservoir contribution does not enter
Eq.~\eqref{eq:app_population2}. Instead, it governs the coherence
$c(t)$ through
\begin{align}
\dot{c}
&=
-\gamma\left(N+\frac{1}{2}\right)c
+
M\gamma'e^{i\phi}c^{*},
\label{eq:app_coherence1}
\\
\dot{c}^{*}
&=
-\gamma\left(N+\frac{1}{2}\right)c^{*}
+
M\gamma'e^{-i\phi}c.
\label{eq:app_coherence2}
\end{align}
Since the battery is initially prepared in $|g\rangle$, one has
$c(0)=0$, and therefore $c(t)=0$ for all times. The single-emitter
state consequently remains diagonal,
\begin{equation}
\rho_1(t)
=
\begin{pmatrix}
p(t) & 0\\
0 & 1-p(t)
\end{pmatrix},
\label{eq:app_single_solution}
\end{equation}
with $p(t)$ given by Eq.~\eqref{eq:app_population_solution}.

Equation~\eqref{eq:app_single_solution} immediately yields the stored
energy reported in the main text,
\begin{equation}
E_1(t)
=
\hbar\omega_0 p(t)
=
\hbar\omega_0
\frac{N}{2N+1}
\left[
1-e^{-\gamma(2N+1)t}
\right].
\label{eq:app_single_energy}
\end{equation}
Moreover, for any finite $N$,
\begin{equation}
p(t)
\leq
\frac{N}{2N+1}
<
\frac{1}{2},
\label{eq:app_population_bound}
\end{equation}
so that $\rho_{gg}(t)>\rho_{ee}(t)$. The larger eigenvalue of
$\rho(t)$ is therefore already associated with the lower-energy state
$|g\rangle$. Hence the state is passive with respect to $H_B$,
implying $E_{\mathrm{pas},1}(t)=E_1(t),$ and consequently
\begin{equation}
\mathcal{W}_1(t)=0.
\label{eq:app_single_ergotropy}
\end{equation}

\section{Two-emitter dynamics}
\label{app:two_emitter}

Here, we derive here the analytical two-emitter solution
used in Sec.~\ref{sec:two}. For two identical emitters, the battery
Hamiltonian is $H_B
=
\hbar\omega_0
\left(
S_1^{+}S_1^{-}
+
S_2^{+}S_2^{-}
\right),$ and the battery is initially prepared in the collective ground state, $\rho_2(0)=|gg\rangle\langle gg|.$

Following Ref.~\cite{you2018waveguide}, we introduce the collective
single-excitation states $|+\rangle
=
\frac{|eg\rangle+|ge\rangle}{\sqrt{2}},$ and $|-\rangle
=
\frac{|eg\rangle-|ge\rangle}{\sqrt{2}},$ and define $\rho_{gg}=\langle gg|\rho|gg\rangle,$ $\rho_{ee}=\langle ee|\rho|ee\rangle,$ $\rho_{++}=\langle +|\rho|+\rangle$ and $\rho_{--}=\langle -|\rho|-\rangle.$ The phase-sensitive two-photon coherence is written as $\rho_u
=
e^{-2ik_{0z}R}\langle ee|\rho|gg\rangle
+
e^{2ik_{0z}R}\langle gg|\rho|ee\rangle.$

The corresponding equations of motion are
\begin{align}
\dot{\rho}_{gg}
={}&
-2N\gamma\rho_{gg}
+(N+1)\gamma_{+}\rho_{++}
+(N+1)\gamma_{-}\rho_{--}
+M\gamma'_{12}\rho_u,
\nonumber\\
\dot{\rho}_{ee}
={}&
-2(N+1)\gamma\rho_{ee}
+N\gamma_{+}\rho_{++}
+N\gamma_{-}\rho_{--}
+M\gamma'_{12}\rho_u,
\nonumber\\
\dot{\rho}_{++}
={}&
-(2N+1)\gamma_{+}\rho_{++}
+(N+1)\gamma_{+}\rho_{ee}
+N\gamma_{+}\rho_{gg}
-M\gamma'_{+}\rho_u,
\nonumber\\
\dot{\rho}_{--}
={}&
-(2N+1)\gamma_{-}\rho_{--}
+(N+1)\gamma_{-}\rho_{ee}
+N\gamma_{-}\rho_{gg}
-M\gamma'_{-}\rho_u,
\nonumber\\
\dot{\rho}_{u}
={}&
-(2N+1)\gamma\rho_u
-2M\gamma'_{+}\rho_{++}
-2M\gamma'_{-}\rho_{--}+
2M\gamma'_{12}
\left(
\rho_{ee}+\rho_{gg}
\right),
\label{eq:app_two_general}
\end{align}
where $\gamma=\gamma_{1d},$ $\gamma_{\pm}
=
\gamma_{1d}
\left[
1\pm\cos(k_{0z}r_{12})
\right],$ $\gamma'_{12}
=
\gamma_{1d}\cos(2k_{0z}r_c)$ and $\gamma'_{\pm}
=
\gamma_{1d}
\left\{
\cos(2k_{0z}r_c)
\pm
\frac{1}{2}
\left[
\cos(2k_{0z}r_1)
+
\cos(2k_{0z}r_2)
\right]
\right\},$ with $r_c=\frac{r_1+r_2}{2}.$

We now specialize to the configuration considered in the main text, $r_c=0$ and $r_{12}=\frac{\lambda_{0z}}{4}.$ Since $k_{0z}r_{12}=\pi/2$, the corresponding coefficients become $\gamma_{+}
=
\gamma_{-}
=
\gamma$ and $\gamma'_{12}
=
\gamma'_{+}
=
\gamma'_{-}
=
\gamma.$

For the initial state, the two single-excitation
populations remain equal, $\rho_{++}(t)=\rho_{--}(t)= P(t).$ We therefore introduce $g=\rho_{gg},$ $e=\rho_{ee}$, $u=\rho_u,$ and $q=2N+1.$

Equation~\eqref{eq:app_two_general} then reduces to
\begin{align}
g'
&=
-2Ng
+
2(N+1)P
+
Mu,
\nonumber\\
e'
&=
-2(N+1)e
+
2NP
+
Mu,
\nonumber\\
P'
&=
-qP
+
(N+1)e
+
Ng
-
Mu,
\nonumber\\
u'
&=
-qu
-
4MP
+
2M(e+g),
\label{eq:app_two_reduced}
\end{align}
where the prime denotes differentiation with respect to $t$. The
initial conditions are $g(0)=1,$ $e(0)=P(0)=u(0)=0,$ and normalization requires $g+e+2P=1.$

Introducing the population difference $d(t)=g(t)-e(t),$ the first two equations in Eq.~\eqref{eq:app_two_reduced} give $d'=1-qd.$ With $d(0)=1$, one obtains
\begin{equation}
d(t)
=
\frac{1+2Ne^{-q\gamma t}}{q}.
\label{eq:app_two_d}
\end{equation}

The single-excitation population follows from $g=\frac{1-2P+d}{2}$ and $e=\frac{1-2P-d}{2}.$ Substituting these expressions into Eq.~\eqref{eq:app_two_reduced}
and eliminating $u$ and $d$ gives
\begin{equation}
P''+3qP'+2P=0,
\end{equation}
with $P(0)=0$ and $P'(0)=N.
$

Defining $\Delta=\sqrt{9q^2-8}$ and $\lambda_{\pm}
=
\frac{-3q\pm\Delta}{2},$ the solution is
\begin{equation}
P(t)
=
\frac{N}{\Delta}
\left(
e^{\lambda_{+}\gamma t}
-
e^{\lambda_{-}\gamma t}
\right).
\label{eq:app_two_p}
\end{equation}
The remaining populations are therefore
\begin{equation}
g(t)
=
\frac{1-2P(t)+d(t)}{2},
\qquad
e(t)
=
\frac{1-2P(t)-d(t)}{2}.
\label{eq:app_two_ge}
\end{equation}

The phase-rotated two-photon coherence follows directly from the
equation for $P$,
\begin{equation}
Mu
=
\frac{q}{2}
-
2qP
-
\frac{d}{2}
-
P',
\end{equation}
and hence
\begin{equation}
u(t)
=
\frac{1}{M}
\Bigg[
\frac{q}{2}
-2q P(t)
-\frac{d(t)}{2}
-
\frac{N}{\Delta}
\left(
\lambda_{+}e^{\lambda_{+}\gamma t}
-
\lambda_{-}e^{\lambda_{-}\gamma t}
\right)
\Bigg].
\label{eq:app_two_u}
\end{equation}
For $N=0$, the continuous vacuum limit of
Eq.~\eqref{eq:app_two_u} gives $u=0$. To reconstruct the corresponding density matrix, we define $\phi=2k_{0z}R$ and $c(t)=\langle ee|\rho(t)|gg\rangle.$

Since $u=e^{-i\phi}c+e^{i\phi}c^{*},$ the phase-rotated quantity $\tilde{c}=e^{-i\phi}c$ remains real for
the present initial condition and dynamics. Thus, $c=\frac{u}{2}e^{i\phi}$ and $c^{*}=\frac{u}{2}e^{-i\phi}.
$
The two-emitter state therefore takes the form
\begin{equation}
\rho_{2}(t)
=
\begin{pmatrix}
g(t) & 0 & 0 & \dfrac{u(t)}{2}e^{-i\phi}\\
0 & P(t) & 0 & 0\\
0 & 0 & P(t) & 0\\
\dfrac{u(t)}{2}e^{i\phi} & 0 & 0 & e(t)
\end{pmatrix},
\label{eq:app_two_rho}
\end{equation}
in the basis
$\{|gg\rangle,|+\rangle,|-\rangle,|ee\rangle\}$.

The stored energy follows from
Eq.~\eqref{eq:app_two_rho} as
\begin{equation}
E_2(t)
=
\hbar\omega_0[1-d(t)],
\end{equation}
which, upon using Eq.~\eqref{eq:app_two_d}, gives
\begin{equation}
E_2(t)
=
\frac{2N}{2N+1}
\left[
1-e^{-(2N+1)\gamma t}
\right]
\hbar\omega_0.
\label{eq:app_two_energy}
\end{equation}

The spectrum of Eq.~\eqref{eq:app_two_rho} consists of the two
eigenvalues $p$ and
\begin{equation}
\Lambda_{\pm}
=
\frac{
1-2P
\pm
\sqrt{d^2+u^2}
}{2}.
\end{equation}
For the ground-state initial condition and the symmetric geometry
considered here, the eigenvalues satisfy
$\Lambda_{+}\geq P\geq\Lambda_{-}$. The passive-state energy is
therefore $E_{\mathrm{pas},2}
=
2\hbar\omega_0
\left(
P+\Lambda_{-}
\right),$ and the ergotropy becomes
\begin{equation}
\mathcal{W}_2(t)
=
\hbar\omega_0
\left[
\sqrt{d^2(t)+u^2(t)}
-
d(t)
\right].
\label{eq:app_two_ergotropy}
\end{equation}

\begin{comment}
Finally, for $r_c=0$ the steady-state solution corresponds to the
$n=0$ case identified in Ref.~\cite{you2018waveguide}. The
populations satisfy $\rho_{gg}^{\mathrm{ss}}
=
\frac{N+1}{2N+1},$ $\rho_{ee}^{\mathrm{ss}}
=
\frac{N}{2N+1}$ and $\rho_{++}^{\mathrm{ss}}
=
\rho_{--}^{\mathrm{ss}}
=
0,$ and the steady state can be written as $|\Psi_{\mathrm{ss}}\rangle
=
\frac{
\sqrt{N+1}\,|gg\rangle
-
e^{i\phi}\sqrt{N}\,|ee\rangle
}
{\sqrt{2N+1}}.$

Because this state is pure, its passive counterpart is the ground
state $|gg\rangle\langle gg|$, so that
\begin{equation}
E_{\mathrm{pas},2}^{\mathrm{ss}}=0.
\end{equation}
Consequently,
\begin{equation}
E_2^{\mathrm{ss}}
=
\mathcal{W}_2^{\mathrm{ss}}
=
\frac{2N}{2N+1}\hbar\omega_0.
\label{eq:app_two_ss_energy}
\end{equation}
\end{comment}

\section{Three-emitter dynamics}
\label{app:three_emitter}

We consider three identical emitters coupled to the one-dimensional
squeezed waveguide introduced solution used in
Sec.~\ref{sec:three_emitter}. For the
geometry used in the main text, the emitter positions are $r_1=-\lambda_{0z},$ $r_2=0$ and $r_3=\lambda_{0z},$ where $\lambda_{0z}=\frac{2\pi}{k_{0z}}$ is the effective longitudinal wavelength of the waveguide. The
corresponding separations are $r_{12}=r_{23}=\lambda_{0z},$ and $r_{13}=2\lambda_{0z}.$ Using $k_{0z}\lambda_{0z}=2\pi$ together with
the emitter positions gives $\gamma_{ij}=\gamma'_{ij}=\gamma_{1d}=\gamma,$ and $\Omega_{ij}=0,$ for the couplings relevant to the three-emitter dynamics. 
%The coherent dipole-dipole contribution therefore vanishes, while both the ordinary collective decay and the phase-sensitive squeezed-reservoir coupling are fully collective.
Introducing the collective operators $J^{+}=\sum_{i=1}^{3}S_i^{+}$ and $J^{-}=\sum_{i=1}^{3}S_i^{-},$
\begin{comment}
the sums appearing in the master equation reduce according to
\begin{subequations}
\begin{align}
\sum_{i,j}S_i^{+}S_j^{-}
&=J^{+}J^{-},
&
\sum_{i,j}S_j^{-}\rho S_i^{+}
&=J^{-}\rho J^{+},
\\
\sum_{i,j}S_i^{-}S_j^{+}
&=J^{-}J^{+},
&
\sum_{i,j}S_j^{+}\rho S_i^{-}
&=J^{+}\rho J^{-},
\\
\sum_{i,j}S_i^{+}S_j^{+}
&=(J^{+})^{2},
&
\sum_{i,j}S_j^{+}\rho S_i^{+}
&=J^{+}\rho J^{+},
\\
\sum_{i,j}S_i^{-}S_j^{-}
&=(J^{-})^{2},
&
\sum_{i,j}S_j^{-}\rho S_i^{-}
&=J^{-}\rho J^{-}.
\end{align}
\end{subequations}    
\end{comment}
the master equation becomes
\begin{align}
\frac{d\rho_3}{dt}
={}&
-\frac{\gamma(N+1)}{2}
\left(
\rho_3 J^{+}J^{-}
+J^{+}J^{-}\rho_3
-2J^{-}\rho_3 J^{+}
\right)
\nonumber\\
&-\frac{\gamma N}{2}
\left(
\rho_3 J^{-}J^{+}
+J^{-}J^{+}\rho_3
-2J^{+}\rho_3 J^{-}
\right)
\nonumber\\
&-\frac{\gamma M}{2}e^{i\phi}
\left[
\rho_3 (J^{+})^{2}
+(J^{+})^{2}\rho_3
-2J^{+}\rho_3 J^{+}
\right]
\nonumber\\
&-\frac{\gamma M}{2}e^{-i\phi}
\left[
\rho_3(J^{-})^{2}
+(J^{-})^{2}\rho_3
-2J^{-}\rho_3 J^{-}
\right],
\label{app:eq:three_collective_ME}
\end{align}

where $\phi=2k_{0z}R$. Because the collective operators act identically on the three
emitters, the evolution preserves exchange symmetry. Starting from
$|ggg\rangle$, the dynamics can therefore be described in the basis
$|0\rangle=|ggg\rangle$,
$|1\rangle=(|egg\rangle+|geg\rangle+|gge\rangle)/\sqrt{3}$,
$|2\rangle=(|eeg\rangle+|ege\rangle+|gee\rangle)/\sqrt{3}$, and
$|3\rangle=|eee\rangle$.

The master equation also preserves excitation parity. Consequently,
coherences between states with opposite parity are not generated from
the initial state, leaving only the
$|0\rangle\leftrightarrow|2\rangle$ and
$|1\rangle\leftrightarrow|3\rangle$ coherence channels. Defining
$P_n=\langle n|\rho|n\rangle$ and absorbing the fixed squeezing phase
into the coherences through
$X=e^{i\phi}\rho_{02}$ and $Y=e^{i\phi}\rho_{13}$, with $X$ and $Y$
real for the present initial condition, the density matrix takes the
form
\begin{equation}
\rho_3(t)=
\begin{pmatrix}
P_0(t) & 0 & e^{-i\phi}X(t) & 0\\
0 & P_1(t) & 0 & e^{-i\phi}Y(t)\\
e^{i\phi}X(t) & 0 & P_2(t) & 0\\
0 & e^{i\phi}Y(t) & 0 & P_3(t)
\end{pmatrix}.
\label{app:eq:three_density_reduced}
\end{equation}

Substitution into Eq.~\eqref{app:eq:three_collective_ME} gives the
closed set of six real equations
\begin{subequations}
\begin{align}
\dot{P}_0
&=
\scalebox{0.82}{$\displaystyle
\gamma\left[
-3NP_0+3(N+1)P_1-2\sqrt{3}MX
\right]
$},
\\
\dot{P}_1
&=
\scalebox{0.82}{$\displaystyle
\gamma\left[
3NP_0-(7N+3)P_1+4(N+1)P_2
+4\sqrt{3}MX-2\sqrt{3}MY
\right]
$},
\\
\dot{P}_2
&=
\scalebox{0.82}{$\displaystyle
\gamma\left[
4NP_1-(7N+4)P_2+3(N+1)P_3
-2\sqrt{3}MX+4\sqrt{3}MY
\right]
$},
\\
\dot{P}_3
&=
\scalebox{0.82}{$\displaystyle
\gamma\left[
3NP_2-3(N+1)P_3-2\sqrt{3}MY
\right]
$},
\\
\dot{X}
&=
\scalebox{0.82}{$\displaystyle
\gamma\left[
-(5N+2)X+3(N+1)Y
-\sqrt{3}M(P_0-2P_1+P_2)
\right]
$},
\\
\dot{Y}
&=
\scalebox{0.82}{$\displaystyle
\gamma\left[
3NX-(5N+3)Y
-\sqrt{3}M(P_1-2P_2+P_3)
\right]
$}.
\end{align}
\label{app:eq:three_EOM}
\end{subequations}
Introducing the vector $\mathbf{v}(t)=
\begin{pmatrix}
P_0&
P_1&
P_2&
P_3&
X&
Y
\end{pmatrix}^{T},$ Eqs.~\eqref{app:eq:three_EOM} can be written as
\begin{equation}
\frac{d\mathbf{v}}{dt}
=
\mathbf{A}\mathbf{v},
\label{app:eq:three_matrix_eom}
\end{equation}
with
\begin{equation}
\mathbf{A}
=
\resizebox{0.9\columnwidth}{!}{$
\begin{pmatrix}
-3N & 3(N+1) & 0 & 0 & -2\sqrt{3}M & 0\\
3N & -(7N+3) & 4(N+1) & 0 & 4\sqrt{3}M & -2\sqrt{3}M\\
0 & 4N & -(7N+4) & 3(N+1) & -2\sqrt{3}M & 4\sqrt{3}M\\
0 & 0 & 3N & -3(N+1) & 0 & -2\sqrt{3}M\\
-\sqrt{3}M & 2\sqrt{3}M & -\sqrt{3}M & 0
& -(5N+2) & 3(N+1)\\
0 & -\sqrt{3}M & 2\sqrt{3}M & -\sqrt{3}M
& 3N & -(5N+3)
\end{pmatrix}
$}.
\label{app:eq:three_matrix_A}
\end{equation}
For the initial ground state $|ggg\rangle$, the corresponding
initial conditions are $P_0(0)=1$ and
$P_1(0)=P_2(0)=P_3(0)=X(0)=Y(0)=0$.
The resulting linear system determines the transient dynamics and its
steady state solution through $\mathbf{A}\mathbf{v}_{\mathrm{ss}}=0$,
together with the normalization
$P_0+P_1+P_2+P_3=1$.

With the state dynamics specified, the energetic properties of the
three-emitter battery can now be evaluated. The battery Hamiltonian is
$H_B=\hbar\omega_0\sum_{i=1}^{3}S_i^{+}S_i^{-}$.
In the full eight-dimensional Hilbert space, the corresponding energy
spectrum is
$\{0,\hbar\omega_0,\hbar\omega_0,\hbar\omega_0,
2\hbar\omega_0,2\hbar\omega_0,2\hbar\omega_0,
3\hbar\omega_0\}$.
Restricting to the symmetric basis
$\{|0\rangle,|1\rangle,|2\rangle,|3\rangle\}$,
the Hamiltonian reduces to
$H_B=\hbar\omega_0\operatorname{diag}(0,1,2,3)$.

Because the state has support only in this subspace, the stored energy
is exactly
\begin{equation}
E_3(t)
=
\hbar\omega_0
\left[
P_1(t)+2P_2(t)+3P_3(t)
\right].
\label{app:eq:E3}
\end{equation}

The ergotropy is determined from the spectrum of the battery state
together with the full energy spectrum of the three-emitter
Hamiltonian. From Eq.~\eqref{app:eq:three_density_reduced}, the density
matrix separates into two independent parity sectors,
$\{|0\rangle,|2\rangle\}$ and $\{|1\rangle,|3\rangle\}$. The four
nonzero eigenvalues are therefore obtained directly from the
corresponding $2\times2$ blocks,
\begin{subequations}
\begin{align}
\lambda_{02,\pm}
&=
\frac{P_0+P_2
\pm\sqrt{(P_0-P_2)^2+4X^2}}{2},
\\
\lambda_{13,\pm}
&=
\frac{P_1+P_3
\pm\sqrt{(P_1-P_3)^2+4Y^2}}{2}.
\end{align}
\label{app:eq:rho3_eigenvalues}
\end{subequations}
Since the dynamics remains confined to the symmetric subspace, the
remaining four eigenvalues of the full eight-dimensional density
operator are zero.
The fixed squeezing phase $\phi$ does not affect these eigenvalues.

The largest eigenvalue of the three-emitter state is obtained by
comparing the upper eigenvalue of the even- and odd-parity sectors,
\begin{equation}
\nu_1(t)
=
\max
\left\{
\lambda_{02,+}(t),
\lambda_{13,+}(t)
\right\}.
\label{app:eq:largest_eigenvalue}
\end{equation}

The passive state is obtained by assigning $\nu_1$ to the unique
ground state and $\nu_2$, $\nu_3$, and $\nu_4$ to the threefold
degenerate one-excitation manifold. The four zero eigenvalues occupy
the remaining higher-energy levels. Hence, the passive energy reduces to $E_{\mathrm{pas},3}(t)
=
\hbar\omega_0
\left[
1-\nu_1(t)
\right].$

Finally,
\begin{equation}
\mathcal{W}_3(t)
=
\hbar\omega_0
\left[
P_1(t)
+2P_2(t)
+3P_3(t)
-1
+\nu_1(t)
\right].
\label{app:eq:W3}
\end{equation}

\bibliography{refs}

@article{you2018waveguide,
  title={Waveguide quantum electrodynamics in squeezed vacuum},
  author={You, Jieyu and Liao, Zeyang and Li, Sheng-Wen and Zubairy, M Suhail},
  journal={Physical Review A},
  volume={97},
  number={2},
  pages={023810},
  year={2018},
  publisher={APS}
}

@article{yang2023battery,
  title={Battery capacity of energy-storing quantum systems},
  author={Yang, Xue and Yang, Yan-Han and Alimuddin, Mir and Salvia, Raffaele and Fei, Shao-Ming and Zhao, Li-Ming and Nimmrichter, Stefan and Luo, Ming-Xing},
  journal={Physical review letters},
  volume={131},
  number={3},
  pages={030402},
  year={2023},
  publisher={APS}
}

@article{barra2019dissipative,
  title={Dissipative charging of a quantum battery},
  author={Barra, Felipe},
  journal={Physical review letters},
  volume={122},
  number={21},
  pages={210601},
  year={2019},
  publisher={APS}
}

@article{campaioli2024colloquium,
  title={Colloquium: quantum batteries},
  author={Campaioli, Francesco and Gherardini, Stefano and Quach, James Q and Polini, Marco and Andolina, Gian Marcello},
  journal={Reviews of Modern Physics},
  volume={96},
  number={3},
  pages={031001},
  year={2024},
  publisher={APS}
}

@article{PhysRevE.102.042111,
  title={Ergotropy from coherences in an open quantum system},
  author={{\c{C}}akmak, Bar{\i}{\c{s}}},
  journal={Physical Review E},
  volume={102},
  number={4},
  pages={042111},
  year={2020},
  publisher={APS}
}

@article{hatsopoulos1976unified,
  title={A unified quantum theory of mechanics and thermodynamics. Part IIa. Available energy},
  author={Hatsopoulos, George N and Gyftopoulos, Elias P},
  journal={Foundations of Physics},
  volume={6},
  number={2},
  pages={127--141},
  year={1976},
  publisher={Springer}
}

@article{allahverdyan2004maximal,
  title={Maximal work extraction from finite quantum systems},
  author={Allahverdyan, Armen E and Balian, Roger and Nieuwenhuizen, Th M},
  journal={EPL (Europhysics Letters)},
  volume={67},
  number={4},
  pages={565--571},
  year={2004}
}

@article{ladd2010quantum,
  title={Quantum computers},
  author={Ladd, Thaddeus D and Jelezko, Fedor and Laflamme, Raymond and Nakamura, Yasunobu and Monroe, Christopher and O'Brien, Jeremy Lloyd},
  journal={nature},
  volume={464},
  number={7285},
  pages={45--53},
  year={2010},
  publisher={Nature Publishing Group UK London}
}

@book{nielsen2010quantum,
  title={Quantum computation and quantum information},
  author={Nielsen, Michael A and Chuang, Isaac L},
  year={2010},
  publisher={Cambridge university press}
}

@article{usenko2026continuous,
  title={Continuous-variable quantum communication},
  author={Usenko, Vladyslav C and Ac{\'\i}n, Antonio and All{\'e}aume, Romain and Andersen, Ulrik L and Diamanti, Eleni and Gehring, Tobias and Hajomer, Adnan AE and Kanitschar, Florian and Pacher, Christoph and Pirandola, Stefano and others},
  journal={Reviews of Modern Physics},
  volume={98},
  number={1},
  pages={015003},
  year={2026},
  publisher={APS}
}

@article{hasan2023quantum,
  title={Quantum communication systems: vision, protocols, applications, and challenges},
  author={Hasan, Syed Rakib and Chowdhury, Mostafa Zaman and Sayem, Mohammad and Jang, Yeong Min},
  journal={IEEE Access},
  volume={11},
  pages={15855--15877},
  year={2023},
  publisher={IEEE}
}

@article{halimeh2025cold,
  title={Cold-atom quantum simulators of gauge theories},
  author={Halimeh, Jad C and Aidelsburger, Monika and Grusdt, Fabian and Hauke, Philipp and Yang, Bing},
  journal={Nature Physics},
  volume={21},
  number={1},
  pages={25--36},
  year={2025},
  publisher={Nature Publishing Group UK London}
}

@article{daley2022practical,
  title={Practical quantum advantage in quantum simulation},
  author={Daley, Andrew J and Bloch, Immanuel and Kokail, Christian and Flannigan, Stuart and Pearson, Natalie and Troyer, Matthias and Zoller, Peter},
  journal={Nature},
  volume={607},
  number={7920},
  pages={667--676},
  year={2022},
  publisher={Nature Publishing Group UK London}
}

@article{karsa2024quantum,
  title={Quantum illumination and quantum radar: A brief overview},
  author={Karsa, Athena and Fletcher, Alasdair and Spedalieri, Gaetana and Pirandola, Stefano},
  journal={Reports on progress in physics},
  volume={87},
  number={9},
  pages={094001},
  year={2024},
  publisher={IOP Publishing}
}

@article{palma2026optimal,
  title={Optimal Quantum Illumination with Nonlocal Non-Gaussian Operations},
  author={Palma, Luis D Zambrano and Maleki, Yusef and Zubairy, M Suhail},
  journal={arXiv preprint arXiv:2605.13747},
  year={2026}
}

@article{pirandola2018advances,
  title={Advances in photonic quantum sensing},
  author={Pirandola, Stefano and Bardhan, B Roy and Gehring, Tobias and Weedbrook, Christian and Lloyd, Seth},
  journal={Nature Photonics},
  volume={12},
  number={12},
  pages={724--733},
  year={2018},
  publisher={Nature Publishing Group UK London}
}

@article{huang2024entanglement,
  title={Entanglement-enhanced quantum metrology: From standard quantum limit to Heisenberg limit},
  author={Huang, Jiahao and Zhuang, Min and Lee, Chaohong},
  journal={Applied Physics Reviews},
  volume={11},
  number={3},
  year={2024},
  publisher={AIP Publishing}
}

@article{4h4b-3xss,
  title = {Information conservation relations for weak measurement and its reversal},
  author = {Maleki, Yusef and Palma, Luis D. Zambrano and Zubairy, M. Suhail},
  journal = {Phys. Rev. A},
  volume = {113},
  issue = {1},
  pages = {012215},
  numpages = {7},
  year = {2026},
  month = {Jan},
  publisher = {American Physical Society},
  doi = {10.1103/4h4b-3xss},
  url = {https://link.aps.org/doi/10.1103/4h4b-3xss}
}

@article{doi:10.1142/S0219477525400280,
author = {Palma, Luis D. Zambrano and Maleki, Yusef and Zubairy, M. Suhail},
title = {Information-Theoretic Analysis of Weak Measurements and Their Reversal},
journal = {Fluctuation and Noise Letters},
volume = {25},
number = {02},
pages = {2540028},
year = {2026},
doi = {10.1142/S0219477525400280},

URL = { 
    
        https://doi.org/10.1142/S0219477525400280
    
    

},
eprint = { 
    
        https://doi.org/10.1142/S0219477525400280
    
    }
}

@article{francica2020quantum,
  title={Quantum coherence and ergotropy},
  author={Francica, Gianluca and Binder, Felix C and Guarnieri, Giacomo and Mitchison, Mark T and Goold, John and Plastina, Francesco},
  journal={Physical Review Letters},
  volume={125},
  number={18},
  pages={180603},
  year={2020},
  publisher={APS}
}

@article{ahmadi2025harnessing,
  title={Harnessing environmental noise for quantum energy storage},
  author={Ahmadi, Borhan and Ravichandran, Aravinth Balaji and Mazurek, Pawe{\l} and Barzanjeh, Shabir and Horodecki, Pawe{\l}},
  journal={arXiv preprint arXiv:2510.06384},
  year={2025}
}

@article{PhysRevLett.113.140401,
  title = {Quantifying Coherence},
  author = {Baumgratz, T. and Cramer, M. and Plenio, M. B.},
  journal = {Phys. Rev. Lett.},
  volume = {113},
  issue = {14},
  pages = {140401},
  numpages = {5},
  year = {2014},
  month = {Sep},
  publisher = {American Physical Society},
  doi = {10.1103/PhysRevLett.113.140401},
  url = {https://link.aps.org/doi/10.1103/PhysRevLett.113.140401}
}

@article{streltsov2017colloquium,
  title={Colloquium: Quantum coherence as a resource},
  author={Streltsov, Alexander and Adesso, Gerardo and Plenio, Martin B},
  journal={Reviews of Modern Physics},
  volume={89},
  number={4},
  pages={041003},
  year={2017},
  publisher={APS}
}

@article{PhysRevE.105.L052101,
  title = {Quantum correlations and ergotropy},
  author = {Francica, Gianluca},
  journal = {Phys. Rev. E},
  volume = {105},
  issue = {5},
  pages = {L052101},
  numpages = {5},
  year = {2022},
  month = {May},
  publisher = {American Physical Society},
  doi = {10.1103/PhysRevE.105.L052101},
  url = {https://link.aps.org/doi/10.1103/PhysRevE.105.L052101}
}

@article{fleischhauer2002quantum,
  title={Quantum memory for photons: Dark-state polaritons},
  author={Fleischhauer, Michael and Lukin, Mikhail D},
  journal={Physical Review A},
  volume={65},
  number={2},
  pages={022314},
  year={2002},
  publisher={APS}
}

@article{lvovsky2009optical,
  title={Optical quantum memory},
  author={Lvovsky, Alexander I and Sanders, Barry C and Tittel, Wolfgang},
  journal={Nature photonics},
  volume={3},
  number={12},
  pages={706--714},
  year={2009},
  publisher={Nature Publishing Group UK London}
}

@article{alicki2013entanglement,
  title={Entanglement boost for extractable work from ensembles of quantum batteries},
  author={Alicki, Robert and Fannes, Mark},
  journal={Physical Review E—Statistical, Nonlinear, and Soft Matter Physics},
  volume={87},
  number={4},
  pages={042123},
  year={2013},
  publisher={APS}
}

@article{ferraro2026opportunities,
  title={Opportunities and challenges of quantum batteries},
  author={Ferraro, Dario and Cavaliere, Fabio and Genoni, Marco G and Benenti, Giuliano and Sassetti, Maura},
  journal={Nature Reviews Physics},
  volume={8},
  number={2},
  pages={115--127},
  year={2026},
  publisher={Nature Publishing Group UK London}
}

@article{kamin2020entanglement,
  title={Entanglement, coherence and charging process of quantum batteries},
  author={Kamin, FH and Tabesh, FT and Salimi, S and Santos, Alan C},
  journal={arXiv preprint arXiv:2006.10476},
  year={2020}
}

@article{hovhannisyan2013entanglement,
  title={Entanglement generation is not necessary for optimal work extraction},
  author={Hovhannisyan, Karen V and Perarnau-Llobet, Mart{\'\i} and Huber, Marcus and Ac{\'\i}n, Antonio},
  journal={Physical review letters},
  volume={111},
  number={24},
  pages={240401},
  year={2013},
  publisher={APS}
}

@article{rossini2019many,
  title={Many-body localized quantum batteries},
  author={Rossini, Davide and Andolina, Gian Marcello and Polini, Marco},
  journal={Physical Review B},
  volume={100},
  number={11},
  pages={115142},
  year={2019},
  publisher={APS}
}

@article{andolina2019extractable,
  title={Extractable work, the role of correlations, and asymptotic freedom in quantum batteries},
  author={Andolina, Gian Marcello and Keck, Maximilian and Mari, Andrea and Campisi, Michele and Giovannetti, Vittorio and Polini, Marco},
  journal={Physical review letters},
  volume={122},
  number={4},
  pages={047702},
  year={2019},
  publisher={APS}
}

@article{centrone2021charging,
  title={Charging batteries with quantum squeezing},
  author={Centrone, Federico and Mancino, Luca and Paternostro, Mauro},
  journal={arXiv preprint arXiv:2106.07899},
  year={2021}
}

@article{li2025enhancing,
  title={Enhancing ergotropy of a quantum battery with coherent chargers: The catalystlike role of indefinite causal order},
  author={Li, Hai and Ma, Haoyuan and Hao, Yaming and Yu, Wenli},
  journal={Physical Review A},
  volume={112},
  number={4},
  pages={042229},
  year={2025},
  publisher={APS}
}

@article{binder2015quantacell,
  title={Quantacell: powerful charging of quantum batteries},
  author={Binder, Felix C and Vinjanampathy, Sai and Modi, Kavan and Goold, John},
  journal={New Journal of Physics},
  volume={17},
  number={7},
  pages={075015},
  year={2015},
  publisher={IOP Publishing}
}

@article{pusz1978passive,
  title={Passive states and KMS states for general quantum systems},
  author={Pusz, Wies{\l}aw and Woronowicz, Stanis{\l}aw L},
  journal={Communications in Mathematical Physics},
  volume={58},
  number={3},
  pages={273--290},
  year={1978},
  publisher={Springer}
}

@article{RevModPhys.93.025005,
  title = {Circuit quantum electrodynamics},
  author = {Blais, Alexandre and Grimsmo, Arne L. and Girvin, S. M. and Wallraff, Andreas},
  journal = {Rev. Mod. Phys.},
  volume = {93},
  issue = {2},
  pages = {025005},
  numpages = {72},
  year = {2021},
  month = {May},
  publisher = {American Physical Society},
  doi = {10.1103/RevModPhys.93.025005},
  url = {https://link.aps.org/doi/10.1103/RevModPhys.93.025005}
}

@article{RevModPhys.89.021001,
  title = {Colloquium: Strongly interacting photons in one-dimensional continuum},
  author = {Roy, Dibyendu and Wilson, C. M. and Firstenberg, Ofer},
  journal = {Rev. Mod. Phys.},
  volume = {89},
  issue = {2},
  pages = {021001},
  numpages = {23},
  year = {2017},
  month = {May},
  publisher = {American Physical Society},
  doi = {10.1103/RevModPhys.89.021001},
  url = {https://link.aps.org/doi/10.1103/RevModPhys.89.021001}
}

@article{RevModPhys.95.015002,
  title = {Waveguide quantum electrodynamics: Collective radiance and photon-photon correlations},
  author = {Sheremet, Alexandra S. and Petrov, Mihail I. and Iorsh, Ivan V. and Poshakinskiy, Alexander V. and Poddubny, Alexander N.},
  journal = {Rev. Mod. Phys.},
  volume = {95},
  issue = {1},
  pages = {015002},
  numpages = {59},
  year = {2023},
  month = {Mar},
  publisher = {American Physical Society},
  doi = {10.1103/RevModPhys.95.015002},
  url = {https://link.aps.org/doi/10.1103/RevModPhys.95.015002}
}

@article{PhysRevA.89.042328,
  title = {Generation, manipulation, and detection of two-qubit entanglement in waveguide QED},
  author = {Gonzalez-Ballestero, C. and Moreno, Esteban and Garcia-Vidal, F. J.},
  journal = {Phys. Rev. A},
  volume = {89},
  issue = {4},
  pages = {042328},
  numpages = {9},
  year = {2014},
  month = {Apr},
  publisher = {American Physical Society},
  doi = {10.1103/PhysRevA.89.042328},
  url = {https://link.aps.org/doi/10.1103/PhysRevA.89.042328}
}

@article{PhysRevX.6.031004,
  title = {Resonance Fluorescence from an Artificial Atom in Squeezed Vacuum},
  author = {Toyli, D. M. and Eddins, A. W. and Boutin, S. and Puri, S. and Hover, D. and Bolkhovsky, V. and Oliver, W. D. and Blais, A. and Siddiqi, I.},
  journal = {Phys. Rev. X},
  volume = {6},
  issue = {3},
  pages = {031004},
  numpages = {13},
  year = {2016},
  month = {Jul},
  publisher = {American Physical Society},
  doi = {10.1103/PhysRevX.6.031004},
  url = {https://link.aps.org/doi/10.1103/PhysRevX.6.031004}
}

@article{PhysRevLett.111.090502,
  title = {Waveguide-QED-Based Photonic Quantum Computation},
  author = {Zheng, Huaixiu and Gauthier, Daniel J. and Baranger, Harold U.},
  journal = {Phys. Rev. Lett.},
  volume = {111},
  issue = {9},
  pages = {090502},
  numpages = {5},
  year = {2013},
  month = {Aug},
  publisher = {American Physical Society},
  doi = {10.1103/PhysRevLett.111.090502},
  url = {https://link.aps.org/doi/10.1103/PhysRevLett.111.090502}
}

@article{b5lx-j66h,
  title = {Quantum battery via reservoir engineering},
  author = {Yang, Jin and Fang, Sibo and Zhao, Chengsong and Shan, Chuanjia and Xiong, Biao},
  journal = {Phys. Rev. Appl.},
  volume = {25},
  issue = {3},
  pages = {034013},
  numpages = {18},
  year = {2026},
  month = {Mar},
  publisher = {American Physical Society},
  doi = {10.1103/b5lx-j66h},
  url = {https://link.aps.org/doi/10.1103/b5lx-j66h}
}

@article{PhysRevLett.132.090401,
  title = {Remote Charging and Degradation Suppression for the Quantum Battery},
  author = {Song, Wan-Lu and Liu, Hai-Bin and Zhou, Bin and Yang, Wan-Li and An, Jun-Hong},
  journal = {Phys. Rev. Lett.},
  volume = {132},
  issue = {9},
  pages = {090401},
  numpages = {8},
  year = {2024},
  month = {Feb},
  publisher = {American Physical Society},
  doi = {10.1103/PhysRevLett.132.090401},
  url = {https://link.aps.org/doi/10.1103/PhysRevLett.132.090401}
}

@article{tirone2025many,
  title={Many-body enhancement of energy storage in a waveguide QED quantum battery},
  author={Tirone, Salvatore and Andolina, Gian Marcello and Calaj{\`o}, Giuseppe and Giovannetti, Vittorio and Rossini, Davide},
  journal={Physical Review A},
  volume={112},
  number={1},
  pages={013717},
  year={2025},
  publisher={APS}
}

@article{PhysRevLett.134.180401,
  title = {Topological Quantum Batteries},
  author = {Lu, Zhi-Guang and Tian, Guoqing and L\"u, Xin-You and Shang, Cheng},
  journal = {Phys. Rev. Lett.},
  volume = {134},
  issue = {18},
  pages = {180401},
  numpages = {8},
  year = {2025},
  month = {May},
  publisher = {American Physical Society},
  doi = {10.1103/PhysRevLett.134.180401},
  url = {https://link.aps.org/doi/10.1103/PhysRevLett.134.180401}
}

@article{43n6-rnj3,
  title = {Giant-Atom Quantum Batteries: Lossless Energy Transfer via Interference Engineering},
  author = {Yan, Ke-Xiong and Liu, Yang and Xiao, Yang and Lin, Jun-Hao and Song, Jie and Chen, Ye-Hong and Nori, Franco and Xia, Yan},
  journal = {Phys. Rev. Lett.},
  volume = {136},
  issue = {18},
  pages = {180401},
  numpages = {7},
  year = {2026},
  month = {May},
  publisher = {American Physical Society},
  doi = {10.1103/43n6-rnj3},
  url = {https://link.aps.org/doi/10.1103/43n6-rnj3}
}

@article{h3tl-lxt8,
  title = {Feedback-enhanced driven-dissipative quantum batteries in waveguide QED systems},
  author = {Yin, Xian-Li and Guo, Meixi and Huang, Jian and Lee, Heung-wing Joseph and Zhang, Guofeng},
  journal = {Phys. Rev. Res.},
  volume = {8},
  issue = {3},
  pages = {L032035},
  numpages = {8},
  year = {2026},
  month = {Aug},
  publisher = {American Physical Society},
  doi = {10.1103/h3tl-lxt8},
  url = {https://link.aps.org/doi/10.1103/h3tl-lxt8}
}

@article{maleki2022distributed,
  title={Distributed phase estimation and networked quantum sensors with W-type quantum probes},
  author={Maleki, Yusef and Zubairy, M Suhail},
  journal={Physical Review A},
  volume={105},
  number={3},
  pages={032428},
  year={2022},
  publisher={APS}
}

@article{maleki2023time,
  title={Time-reversal-symmetry breaking in a scalable cavity QED lattice},
  author={Maleki, Yusef and Zhou, Chaofan and Zubairy, M Suhail},
  journal={Physical Review A},
  volume={108},
  number={6},
  pages={063709},
  year={2023},
  publisher={APS}
}

@article{zhou2023emission,
  title={Emission of an atom inside a one-dimensional atomic cavity},
  author={Zhou, Chaofan and Maleki, Yusef and Liao, Zeyang and Zubairy, M Suhail},
  journal={Physical Review A},
  volume={108},
  number={1},
  pages={013708},
  year={2023},
  publisher={APS}
}

\end{document}